\documentclass[reprint,aps,prd]{revtex4-1}
\usepackage[utf8]{inputenc}
\usepackage[utf8]{inputenc}
\usepackage{amsmath}
\usepackage{amsfonts}
\usepackage{amssymb}
\usepackage[left=2cm,right=2cm,top=2cm,bottom=2cm]{geometry}
\usepackage{braket}
\usepackage{dsfont}

\DeclareMathOperator{\tr}{\operatorname{tr}}
\DeclareMathOperator{\MPS}{\operatorname{MPS}}
\DeclareMathOperator{\M}{\textbf{M}}

\usepackage{graphicx}
\usepackage{tikz}
\usetikzlibrary{arrows,decorations.pathmorphing,backgrounds,positioning,fit,petri,shapes,snakes}

\begin{document}

\title{Topological Defects in Quantum Field Theory with Matrix Product States}

\begin{abstract}
Topological defects (kinks) in a relativistic $\phi^{4}$ scalar field theory in $D=(1+1)$ are studied using the matrix product state tensor network. The one kink state is approximated as a matrix product state and the kink mass is calculated. The approach used is quite general and can be applied to a variety of theories and tensor networks. Additionally, the contribution of kink-antikink excitations to the ground state is examined and a general method to estimate the scalar mass from equal time ground state observables is provided. The scalar and kink mass are compared at strong coupling and behave as expected from universality arguments. This suggests that the matrix product state can adequately capture the physics of defect-antidefect excitations and thus provides a promising technique to study challenging non-equilibrium physics such as the Kibble-Zurek mechanism of defect formation.
\end{abstract}

\author{Edward Gillman}
\email{eg909@ic.ac.uk}
\author{Arttu Rajantie}
\email{a.rajantie@imperial.ac.uk}
\affiliation{Department of Physics, Imperial College London, SW7 2AZ, UK}
\date{31 October 2017}
\maketitle

\section{Introduction : Topological Defects and Tensor Networks}

A significant problem within quantum field theory (QFT) is the calculation of non-perturbative and non-equilibrium problems. Standard perturbation theory can be used for near-equilibrium weak coupling problems but fails in other cases. For equilibrium calculations, lattice theory provides a powerful approach but has limited applicability to non-equilibrium problems \cite{Smit2002}. On the non-equilibrium side, the use of \textit{2-particle irreducible} (2PI) or more generally nPI effective actions is common \cite{Berges2004} and can be used in conjunction with large $N$ expansions for a power series approach while \textit{stochastic quantisation} allows for the extension of non-perturbative lattice techniques to the case of real-time \cite{Berges2005}. Recently, \textit{Hamiltonian truncation} techniques have also emerged as an alternative to the available functional/lattice techniques and have been applied to both equilibrium \cite{Rychkov2015} and non-equilibrium calculations \cite{Rakovszky2016}. 

A particularly good test for the various available techniques is offered by the study of \textit{topological defects} in QFT. Topological defects are naturally non-perturbative and cannot be included by perturbative expansions about classical vacuum solutions \cite{Rajaraman1987}. Instead, the QFT is split into different topological sectors, associated to a particular topological charge, and each sector must be treated separately in perturbative expansions. For non-perturbative equilibrium calculations involving defects, lattice techniques have been successful \cite{Groeneveld1981,Rajantie2009,Rajantie2010,Rajantie2012} and more recently Hamiltonian truncation has also been used \cite{Bajnok2015}. However, non-equilibrium calculations can be more problematic as highlighted in the study of defect formation during phase transitions. Despite having a simple and general description via the Kibble-Zurek mechanism \cite{Rajantie2002}, naive applications of the $2PI$ effective action techniques fail to include the contribution of defects \cite{Rajantie2006} requiring the use of less standard techniques \cite{Berges2011}.  Additionally, simple observables such as the quasi-particle excitation density in general fail to capture the relevant physics in such problems \cite{Uhlmann2010} .

In this paper we apply a recent approach based on the use of \textit{matrix product states} (MPS) to the study of topological defects. These techniques deal with states directly and are most commonly used to approximate the ground states of strongly-coupled quantum systems in $D=(1+1)$. While originally specialised to the study of gapped, finite size lattice systems with open boundary conditions \cite{Schollwock2011}, these techniques have seen significant developments in recent years and have been applied to e.g. periodic systems \cite{Verstraete2004,Pippan2010,Rossini2011}, infinite volume systems \cite{Haegeman2013, Zauner-Stauber2017} and non-equilbrium calculations \cite{Haegeman2011,Wall2012}. More generally, matrix product states are part of a broader class of \textit{tensor networks} (TN) which allow further extensions to e.g. critical/gapless systems \cite{Vidal2006,Evenbly2014} and the equilbrium study of excited states (including kink excitations) \cite{Haegeman2012,Porras2006}. Matrix product states have also been applied previously to the study of field theories in a number of settings. For example, they have been applied to lattice regularised relativistic field theories in $D = (1+1)$ with both global symmetries \cite{Weir2010,Milsted2013} and gauge symmetries with substantial progress being made in the Schwinger model ($U(1)$) \cite{Banuls2013,Rico2014,Buyens2014,Pichler2016,Buyens2016}, including an explcit example overcoming the sign problem \cite{Banuls2017}, in addition to studies in the $SU(2)$ gauge theory \cite{Kuhn2015}. More recently, there has also been a focus on the description of field theories using MPS and TN in the continuum, providing an alternative regularisation scheme to the lattice \cite{Ganahl2016,Verstraete2010,Jennings2015}. Finally, while many of the studies involving TN have been for $D = (1+1)$, there have been a number of recent developments towards the application of tensor network methods to higher dimensions, particularly to $D = (2+1)$ using the projected-entangled-pairs states (PEPS) TN \cite{Vanderstraeten2016,Lubasch2014}. These developments include work towards the application of TN to gauge theories in $D=(2+1)$, \cite{Zapp2017}, making TN a promising tool for the future study of high energy physics.

In the following, we apply matrix product states to the study of topological defects (kinks) in the $\phi^{4}$ theory in $D=(1+1)$. We study the equilibrium physics by obtaining MPS approximations to the ground state and one kink state in the lattice regularised setting. The main goal is to assess to what degree the MPS can capture the physics of defects non-perturbatively, both in the sense of providing a direct approximation to the one kink state and in capturing the contribution of kink-antikink excitations to the ground state observables. To achieve this, we focus on studying the kink mass $M_{K}$ and the scalar mass $m_{S}$ at strong and weak couplings where we can compare with analytic results. Confirming that the MPS can capture the physics of defects, particularly of the kink-antikink excitations, is essential if they are to be used to study non-equilibrium phenomena such as defect formation. 

The structure of our discussion will be as follows : Firstly, we will introduce the $\phi^{4}$ theory in Section \ref{phi4} and review some weak coupling and strong coupling results before introducing the lattice regularisation used throughout. Secondly, we review the construction of matrix product states in Section \ref{MPS} and discuss their relationship to entanglement and tensor networks more generally. Following this, we provide an overview of the standard methods used to find MPS approximations to the ground state and show how they apply directly to one kink state in Section \ref{Min}. More details on the information outlined in Sections \ref{MPS} and \ref{Min} , along with a useful guide to implementation,  can be found in the review \cite{Schollwock2011}. In Section \ref{Scalar} we provide a discussion of how to estimate the scalar mass from the equal time two point functions before examining some weak and strong coupling results in Section \ref{Res} and concluding in Section \ref{Concl}. 

\section{$\phi^{4}$ Scalar Field Theory in $D = (1+1)$}
\label{phi4}

\subsection{Classical and Semi-Classical Results}
\label{phi4_SC}

In this section, we review some classical and semi-classical results for the $\phi^{4}$ scalar field theory in $D = (1+1)$ as defined by the action 

\begin{align}
S[\phi] = \int dx dt \left[ \tfrac{1}{2}(\partial_{t}\phi)^{2} - \tfrac{1}{2}(\partial_{x}\phi)^{2} - \tfrac{\mu_{0}^{2}}{2} \phi^{2} -\tfrac{\lambda_{0}}{4!} \phi^{4} \right] ~ . 
\end{align}
When $\mu_{0}^{2} < 0$, the classical potential density
\begin{align}
U(\phi) = \frac{\mu_{0}^{2}}{2} \phi^{2} + \frac{\lambda_{0}}{4!} \phi^{4}
\end{align}
has two degenerate minima (vacua) $\pm v = \pm \sqrt{\tfrac{-6 \mu_{0}^{2}}{\lambda_{0}}}$ corresponding to the spontaneous breaking of the global $\mathds{Z}_{2}$ symmetry $\phi(x) \to -\phi(x)$. Such static, uniform field configurations solve the classical equations of motion but there are additional topologically non-trivial static solutions called kinks $\phi_{K}(x)$ given by
\begin{align}
\phi_{K}(x) = v \tanh \left( \frac{\mu_{0} \left( x - x_{0} \right)}{\sqrt{2}} \right) ~ 
\end{align}
with the corresponding antikink given by $-\phi_{K}(x)$. Such solutions interpolate between the two minima and are not spatially uniform. Additionally, they have a degree of freedom (zero-mode) $x_{0}$ such that they form a continuous family of solutions. These kink solutions are the simplest possible example of a topological defect in quantum field theory. To relate these solutions to a topological charge, all the finite energy field configurations of the theory can be classified according to their homotopy. Configurations in distinct homotopy classes cannot be continuously deformed into one another (e.g. by time evolution) and the theory is split into distinct topological sectors. These sectors are labelled by a topological charge $Q$ determined by the boundary conditions of the configurations as
\begin{align}
Q = \frac{1}{2 v} \left( \phi(\infty) - \phi(-\infty) \right) ~ .
\end{align}
The minima $\pm v$ have $Q=0$ while the kink and antikink have $Q = 1 , -1$ respectively and they provide the lowest energy configurations for each of these sectors. The existence of distinct topological sectors can be confirmed from a few general features of a theory so that one can check easily for the possibility of topological defects in a wide variety of cases \cite{Vachaspati2006}. 

The classical kink mass $M_{K}$ can be calculated from the classical energy
\begin{align}
E = \int dx \left[ \tfrac{1}{2} \left( \partial \phi/\partial x \right)^{2} + U(\phi) \right] 
\end{align}
by subtracting the classical vacuum energy to give
\begin{align}
M_{K} = 4 \sqrt{2} \frac{\mu_{0}^{3}}{\lambda_{0}} ~~ . 
\end{align}

In the quantum theory, the kink and anti-kink will appear as charged particles. Since they lie outside the $Q=0$ vacuum sector, they are non-perturbative in the sense one cannot include their contribution by using perturbation theory starting from the vacuum sector. Instead, to gain semi-classical information about the defects, one must begin from the appropriate sector e.g. with classical kink configurations. Results in these sectors tends to require more work than the topologically trivial $Q=0$ sector  both due to the fact the classical kink is not spatially uniform and due to the presence of the zero-mode $x_{0}$. 

The one-loop order calculation of the kink mass is a classic result known as the `DHN' result following the work of Dashen, Hasslacher and Neveu  \cite{Dashen1974_1,Dashen1974_2}. This can be written in terms of the scalar mass $m_{S}^{2} = 2 \mu_{0}^{2} + \mathcal{O}(\lambda_{0})$ up to $\mathcal{O}(\lambda_{0})$ to give
\begin{align}
M_{K} = 2 \frac{m_{S}^{3}}{\lambda_{0}} + \frac{m_{S}}{2} \left( \frac{1}{6} \sqrt{\frac{3}{2}} - \frac{3}{\pi \sqrt{2}} \right) + \mathcal{O}\left( \lambda_{0} \right) ~.
\end{align}
From this expression one can see that in the semi-classical region the kink appears as a heavy particle such that $M_{K} = \mathcal{O}(m_{S}^{3})$. As such, in this region kink-antikink excitations will provide a negligible contribution to ground state observables. More recently,  zeta-function regularisation has allowed for one-loop results in finite size systems \cite{Pawellek2009} and dimensional regularisation has provided a more systematic approach to one-loop corrections allowing for results at finite temperatures and in higher dimensions \cite{Rebhan2009}. While giving equivalent results to one-loop order, a rigorous treatment of the zero-mode requires more work through e.g. the use of canonical coordinates \cite{Christ1975}, which also allow for the computation of the scalar field $n$-point functions in the presence of the kink \cite{Gervais1975,Papageorgakis2014}. 

While semi-classical expansions can provide information about the weak-coupling region $\mu_{0}^{2} \gg \lambda_{0}$, at stronger couplings they will breakdown. This is particularly important for the $\phi^{4}$ theory in $D = (1+1)$ since the phase transition corresponds to strong couplings. Therefore, near the critical point perturbation theory can no longer be used and one must turn to alternative methods. 

\subsection{Strong-Coupling Results : Universality}
\label{phi4_U}

On approach to the critical point, the correlation length of the $\phi^{4}$ theory diverges. In this regime, the microscopic (lattice) details are irrelevant and the critical behaviour is described by a simple field theory. The nature of this critical field theory depends only a few general features of the underlying theory, e.g. symmetry or dimensionality, so that many different theories have the same critical description and can be separated into universality classes. In $D=2$ the $\phi^{4}$ theory falls into the same universality class as the classical $D=2$ Ising model. The critical field theory can be determined by the fixed points of a suitable renormalisation group (RG) flow and in this case corresponds to a Wilson-Fisher fixed point \cite{Peskin1995}. Since the corresponding critical field theory is not a free scalar field theory, standard perturbation theory cannot be used and we can say the phase transition is strongly coupled. Analytic results can instead be gained for the universality class by using RG or studying the Ising Model which is integrable.  The most familiar universal results are the critical exponents and for the $D=2$ Ising universality class the correlation length $\xi$ in the symmetric and symmetry broken phases is given by 
\begin{align}
\xi &\approx \xi_{0} |\tau|^{\nu}  ~~ \tau >0  \\
\xi &\approx \xi'_{0} |\tau|^{\nu'} ~ \tau < 0  
\end{align}
where $\tau$ is the reduced temperature such that $\tau >0$ indicates the symmetric (high temperature) phase , $\tau < 0$ the symmetry broken (low temperature) phase and $\tau = 0$ the critical point. Additionally, \textit{universal amplitude ratios} \cite{Delfino1998} can be derived and in particular 
\begin{align}
\xi_{0}/\xi'_{0} &\approx 2 ~.
\label{UAR}
\end{align}
These results can be related to the topological defects in the $\phi^{4}$ theory through the Kramers-Wannier duality present in the Ising Model. While explicit kink creation operators cannot be constructed for the $\phi^{4}$ theory, the corresponding disorder operators $\mu(x)$ can be introduced in the classical Ising model \cite{Fradkin2017}. The Kramers-Wannier duality relates the disorder operator two point correlation function on the dual lattice at coupling (temperature) $K$ to the spin operator two point function at a coupling $K^{*}$ as
\begin{align}
\braket{\mu(x) \mu(x')}_{K} = \braket{\sigma(\tilde{x}) \sigma(\tilde{x}')}_{K^{*}} .
\end{align}
This duality establishes the relation $\mu' = \nu$ between the critical exponents where $\mu'$ is the critical exponent associated to the diverging correlation length $\xi_{K}$ of the disorder two point function $\braket{\mu(x) \mu(x')}$ in the symmetry broken phase. When combined with the universal amplitude ratio Equation (\ref{UAR}) one can establish 
\begin{align}
\xi_{K}/\xi &  \approx \frac{\xi_{0} |\tau|^{\nu} }{\xi_{0}'|\tau|^{\nu'}} \nonumber \\
&\approx 2
\end{align}
which uses the hyperscaling relation $\nu = \nu'$. This result corresponds in the $\phi^{4}$ theory to the relationship $m_{S} \approx 2 M_{K}$ between the scalar mass $m_{S}$ and the kink mass $M_{K}$. While universality establishes this result rigorously in the vicinity of the critical point, physically it should hold from the point where first $m_{S} \approx 2 M_{K}$ down to the critical point, since in this region the scalar excitation will decay into a kink-antikink pair excitation which is the lightest excitation for the $Q=0$ sector in this region. Since this result incorporates a simple linear scaling $\nu = 1$ with strong coupling physics and topological defects it provides a good test for non-perturbative calculation methods.

\subsection{Lattice Methods for Non-Perturbative Calculations}

For general non-perturbative calculations it is common to use lattice methods. Essentially, one approximates observables of the full theory by observables of a theory defined on a lattice. An appropriate lattice theory can be obtained for a particular QFT by discretising the continuum action. Usually, this discretisation takes place in both time and space with Euclidean spacetime being used to allow for sampling, effectively transforming the problem of quantum field theory into statistical mechanics. Since analytically continuing back to real time requires further assumptions in the discrete case and tends to dramatically increase the errors from sampling, this method is essentially restricted to equilibrium physics and has been highly successful in this regard, for an introduction to these techniques see \cite{Smit2002}. Here, we will be interested in working with states at a particular time so that the lattice discretisation need only be applied in the spatial dimension leaving time continuous. In this case, the full QFT Hilbert space is truncated down to a lattice quantum theory and we can write schematically $\mathcal{H}_{QFT} \to \mathcal{H}(a,L)$ where the lattice spacing $a$ and size $L$ act as truncation parameters. For spatial discretisation, the Hamiltonian formalism can be used and the continuum theory can be written as
\begin{align}
H[\phi] = \int dx \left[ \tfrac{1}{2}(\partial_{t}\phi)^{2} + \tfrac{1}{2}(\partial_{x} \phi )^{2} + \tfrac{\mu_{0}^{2}}{2} \phi^{2} + \tfrac{\lambda_{0}}{4!} \phi^{4} \right] ~ . 
\label{HamCont}
\end{align}
An appropriate lattice theory can then be constructed by discretisation of this Hamiltonian. A good option is to simply replace the spatial derivatives by first order finite difference approximations leading to the Hamiltonian
\begin{align}
\tilde{H}[\phi] = \sum_{x} \left[ \tfrac{1}{2}(\pi_{x})^{2} + \tfrac{1}{2}(\phi_{x+a} - \phi_{x} )^{2} + \tfrac{\tilde{\mu}_{0}^{2}}{2} \phi_{x}^{2} + \tfrac{\tilde{\lambda}_{0}}{4!} \phi_{x}^{4} \right] 
\label{Ham}
\end{align}
which is given in lattice units with $\tilde{H} = a H , \tilde{\mu}_{0}^{2} = a^{2} \mu_{0}^{2} , \tilde{\lambda}_{0} =  a^{2} \lambda_{0} $. Additionally, the time-derivatives have been replaced by $\pi_{x} = \partial \mathcal{L}/\partial{(\partial_{t}\phi)} = a (\partial_{t}\phi) $ which obeys the canonical commutation relation $[\pi_{x},\phi_{y}] = i \delta_{x,y}$. The finite lattice spacing, which we will set to one throughout, serves to restrict the possible momentum modes $-\pi/a \le p \le \pi/a$ so that low momentum (long distance) observables of the full theory can be well approximated in the lattice theory, while higher momentum (short distance) behaviour will be modified significantly by the lattice. We will be particularly interested in the equal-time observables of the ground state $\braket{\Omega|\mathcal{O}[\phi]|\Omega}$ and the one kink state $\braket{K|\mathcal{O}[\phi]|K}$. These are approximated in the lattice theory by the corresponding lattice observables $\braket{\Omega(a,L)|\mathcal{O}[\phi]|\Omega(a,L)}$ and $\braket{K(a,L)|\mathcal{O}[\phi]|K(a,L)}$ where the state $\ket{\Omega(a,L)}$ is simply the minimum energy state of the lattice theory and $\ket{K(a,L)}$ is the lattice one kink state. To approximate the one kink state on the lattice it is typical to write it explicitly as the minimum energy state of the $Q=1$ topological sector which can be selected out by enforcing twisted periodic boundary conditions (TPBC) in the spatial dimension of the theory $\phi_{x + L} = -\phi_{x}$ \cite{Groeneveld1981}. Both states of interest can then be written in terms of energy minimisation problems as 
\begin{align}
\ket{\Omega(a,L)} = \min_{\ket{\psi}} \left( \braket{\psi|\tilde{H}|\psi} - \lambda \left[\braket{\psi|\psi} - 1 \right] \right) ~
\label{MinGS}
\end{align}
for the lattice ground state and 
\begin{align}
\ket{K(a,L)} = \min_{\ket{\psi}} \left( \braket{\psi|\tilde{H}_{\scriptscriptstyle{(TPBC)}}|\psi} - \lambda \left[\braket{\psi|\psi} - 1\right] \right) 
\label{MinK}
\end{align}
for the one kink lattice state (note that higher kink number excitations are suppressed exponentially with the lattice size $L$). Specifying states further requires the choice of a basis. The field eigenbasis is a natural choice for mean field theory type approximations but to go beyond these it is better to pick a numerically stable basis of real space harmonic oscillators \cite{Drell1976_1}. Introducing real space creation and annihilation operators via  $\phi_{x}= \tfrac{1}{\sqrt{2}} \left( a^{\dag}_{x}+ a_{x}\right)$ and $[a_{x} , a^{\dag}_{y}] = \delta_{x,y}$,  a natural basis set is then the tensor products $\ket{\mathbf{n_{x}}} = \ket{n_{1}}\otimes\ket{n_{2}}\otimes...\otimes\ket{n_{L}}$ where $\ket{n_{x}}$ are the eigenstates of the number operator $N_{x} = a^{\dag}_{x}a_{x}$ at each site. All basis states are now labelled by a $L$-tuple $\mathbf{n_{x}} = ( n_{1} , n_{2} , n_{3} , ... ,n_{L})$ with $ n_{x} \in \mathds{Z}$ and a general state $\ket{\psi}$ can be written as
\begin{align}
\ket{\psi} &= \sum_{n_{1}=0}^{\infty} ... \sum_{n_{L}=0}^{\infty}  \braket{ n_{1} n_{2} ... n_{L} | \psi} \ket{ n_{1} n_{2} ... n_{L}} \nonumber \\
& = \sum_{\mathbf{n_{x}}} \psi_{\mathbf{n_{x}}} \ket{\mathbf{n_{x}}} 
\label{State}
\end{align}
where the state coefficient $\psi_{\mathbf{n_{x}}}$ (wavefunction) now specifies the state in this basis. 

\section{Tensor Networks and Matrix Product States}
\label{MPS}

\subsection{Matrix Product States}

To study the ground state and one kink state observables in the $\phi^{4}$ theory, we would like to solve the minimisation problems Equations (\ref{MinGS}) and (\ref{MinK}) directly. Of course, this is not possible in general since even a finite dimensional Hilbert space grows exponentially with the number of lattice sites. However, matrix product states provide a method to truncate the Hilbert space further down to a tiny subset where states can be specified \textit{efficiently} i.e. with a cost that rises at most polynomially in the number of lattice sites. Within this subset, the minimisation problems can be solved numerically and direct approximations of the ground state and one kink state can be obtained.

To proceed, one first rewrites the expression for a generic state in the theory Equation (\ref{State}) in the matrix product form. For the lattice MPS used here, this first requires an additional truncation in the local Hilbert space dimension such that the resulting total state space is finite. This can be achieved by simply keeping the first $n = 0 , 1 , ... , n_{max} $ basis states at each site. A state can then be expressed as
\begin{align}
\ket{\psi} &= \sum_{n_{1}=0}^{n_{max}} ... \sum_{n_{L}=0}^{n_{max}}  \braket{ n_{1} n_{2} ... n_{L} | \psi} \ket{ n_{1} n_{2} ... n_{L}} ~ . 
\end{align}
When using the field eigenbasis, the error associated with a similar truncation can be rigourously bound by the magnitude of local expectation values $\braket{\psi | \phi_{x}^{2}|\psi}$ and $\braket{\psi | \pi_{x}^{2}|\psi}$ or alternatively the energy expectation value $ E = \braket{\psi | H | \psi }$ \cite{Jordan2014}. In practice, since the limit $ n_{max} \to \infty$ tends to be well behaved, it is usually possible to simply increase the value of $n_{max}$ in calculations until convergence in the desired observables is reached. To keep notation standard, we will let $n_{max} = d-1$ so that the total dimension of the regularised theory is now given by $d^{L}$. The matrix product form can now be introduced by considering the wavefunction $\psi_{\mathbf{n_{x}}}$ as a rank-$L$ tensor. Such a tensor can always be decomposed using a \textit{tensor train decomposition} into a matrix product state form so that
\begin{align}
\psi_{\mathbf{n_{x}}} &= \sum_{\alpha_{1},\alpha_{2},..,\alpha_{L}} M^{n_{1}}_{\alpha_{1},\alpha_{2}}(1) M^{n_{2}}_{\alpha_{2},\alpha_{3}}(2) ...  M^{n_{L}}_{\alpha_{L},\alpha_{1}}(L) \nonumber \\
& = \tr \left(\M^{n_{1}}(1) \M^{n_{2}}(2) ... \M^{n_{L}}(L)\right) \nonumber \\
& = \tr \left( \prod_{x}  \M^{n_{x}}(x) \right)  ~ . 
\end{align}
Here, the state coefficient $\psi_{\mathbf{n_{x}}}$ has been re-expressed as the nearest-neighbour contraction of a set of rank-$3$ tensors, one per site. The size of the tensors can vary site-to-site but we fix them for simplicity to all be $(d,\chi,\chi)$. The first index provides the \textit{physical index} corresponding to the local basis state $\ket{n_{x}}  = \ket{0},\ket{1},...,\ket{d-1}$ while the latter two provide the \textit{internal} or \textit{virtual} degrees of freedom. All the internal indices are contracted over while the physical indices remain uncontracted. The MPS form is made more intuitive by the convenient diagrammatic notation available for tensor networks. In this notation, tensors are represented by shapes with legs corresponding to the indices of the tensor. Contractions between indices are then indicated by the joining of two legs. For example, the rank-3 tensor $M^{i}_{\alpha,\beta}$ is represented by

\begin{center}
\begin{tikzpicture}
[A/.style = {shape=circle,draw=blue!100,fill=gray!50},empty/.style = {shape=circle,draw=blue!0},node distance = 10pt]
\node (M1) at (0,0) [A] {};
\draw [-] (M1) to (0,-0.5);
\draw [-] (M1) -- (-0.5,0);
\draw [-] (M1) -- (0.5,0);
\node[empty] (L1) [ left=of M1  ] {$M^{i}_{\alpha , \beta} = $}; 
\node[empty] (R1) [ right=of M1  ] {$ $}; 
\end{tikzpicture}
\end{center}
such that the PBC lattice MPS with $L = 7$ sites is represented by

\begin{center}
\begin{tikzpicture}
[node distance = 10pt ,A/.style = {shape=circle,draw=blue!100,fill=gray!50} , empty/.style = {shape=circle,draw=blue!0}];

\node[A] (M3) at (0,0) {} ;
\node[A] (M2) [left=of M3] {};
\node[A] (M1) [left=of M2] {};
\node[A] (M4) [right=of M3] {};
\node[A] (M5) [right=of M4] {};
\node[A] (BL) [left=of M1] {};
\node[A] (BR) [right=of M5] {};

\node[empty] (M1B) [below=of M1] {};
\node[empty] (M2B) [below=of M2] {};
\node[empty] (M3B) [below=of M3] {};
\node[empty] (M4B) [below=of M4] {};
\node[empty] (M5B) [below=of M5] {};

\node[empty] (BLB) [below=of BL] {};
\node[empty] (BRB) [below=of BR] {};

\node[circle split] (M1A) [above=of M1] {};
\node[empty] (M2A) [above=of M2] {};
\node[empty] (M3A) [above=of M3] {};
\node[empty] (M4A) [above=of M4] {};
\node[circle split] (M5A) [above=of M5] {};

\node[empty] (L1) [left = of BL] { $\psi_{\textbf{n}} = $};

\draw (M1) -- (M2);
\draw (M2) -- (M3);
\draw ( M3) -- (M4);
\draw (M4) -- (M5);

\draw (M1) -- (M1B);
\draw (M2) -- (M2B);
\draw ( M3) -- (M3B);
\draw (M4) -- (M4B);
\draw (M5) -- (M5B);

\draw (M1) -- (BL);
\draw (M5) -- (BR);

\draw (BLB) -- (BL);
\draw (BRB) -- (BR);

\draw (BL) to  [in=180,out=180] (M1A);

\draw (M1A) to (M5A);

\draw (BR) to  [in=0,out=0] (M5A);

\node[empty] (R1) [ right=of BR ] {$.$}; 

\end{tikzpicture}

\end{center}
Often, as in the original DMRG algorithm, MPS with open boundary conditions (OBC) are used. These can be written by considering the first and last tensors as rank-$2$ tensors so that 
\begin{align}
\psi_{\mathbf{n_{x}}}  = (m_{L}^{n_{1}})^{T} \left(\prod_{x = 2}^{L-1}  \M^{n_{x}}(x) \right) m_{R}^{n_{L}} 
\end{align}
where $m_{L}^{n_{1}}, m_{R}^{n_{L}}$ are rank-$2$ tensors of size $(d,\chi)$ and the trace is no longer needed. The diagrams for OBC MPS are somewhat simpler than their PBC counterparts e.g. for $L=5$
\begin{center}
\begin{tikzpicture}
[node distance = 10pt ,A/.style = {shape=circle,draw=blue!100,fill=gray!50} , empty/.style = {shape=circle,draw=blue!0}];

\node[A] (M3) at (0,0) {} ;
\node[A] (M2) [left=of M3] {};
\node[A] (M1) [left=of M2] {};
\node[A] (M4) [right=of M3] {};
\node[A] (M5) [right=of M4] {};

\node[empty] (M1B) [below=of M1] {};
\node[empty] (M2B) [below=of M2] {};
\node[empty] (M3B) [below=of M3] {};
\node[empty] (M4B) [below=of M4] {};
\node[empty] (M5B) [below=of M5] {};

\node[empty] (L1) [left = of M1] { $\psi_{\textbf{n}} = $};

\draw (M1) -- (M2);
\draw (M2) -- (M3);
\draw ( M3) -- (M4);
\draw (M4) -- (M5);

\draw (M1) -- (M1B);
\draw (M2) -- (M2B);
\draw ( M3) -- (M3B);
\draw (M4) -- (M4B);
\draw (M5) -- (M5B);

\node[empty] (R1) [ right=of M5  ] {$.$}; 
\end{tikzpicture}
\end{center}
Due to this simplification, for convenience we will often use OBC in diagrams, though calculations will tend to involve PBC.

We emphasise that this MPS form is \textit{complete} in the sense that all states can be expressed in this manner. However, the real power of MPS is revealed by considering the subset of states that can be efficiently parametrised by the MPS form. The number of parameters required to specify a MPS is simply given by the lattice size multiplied by the size of the tensor i.e $ Ld\chi^{2}$. A general state requires an exponentially large number of parameters to specify so that at most $\chi \sim d^{L/2}$. However, some states, often simply called ``matrix product states" require only $\chi \sim const$ and these states make up the natural efficiently parametrised subset of the MPS form. By working with the MPS form and picking some value of $\chi$, we can truncate the Hilbert space sufficiently so that we can deal with states directly, i.e. without the need for sampling. This allows one to approximate the states of interest by solving minimisation problems within this subset. Schematically, the series of truncations can then be shown as $\mathcal{H}_{QFT} \to \mathcal{H}(a,L) \to \MPS(a,L,d,\chi)$ where the subset of states $\MPS(a,L,d,\chi)$ no longer retains the structure of $\mathcal{H}$ but can be shown in many cases to form a smooth manifold, for details see \cite{Haegeman2014}.

While states in the subset $\MPS(a,L,d,\chi)$ can be represented efficiently by a tensor network, the central question is which states actually belong to the subset and what physics is well approximated within it. The answer to this is given by considering the entanglement entropy of a reduced state $\rho_{\mathcal{A}}$ for a state in the subset, defined over a spatial subregion $\mathcal{A}$. The entanglement entropy is then defined as
\begin{align}
S(\rho_{\mathcal{A}}) = - \tr\left( \rho_{\mathcal{A}} \log \rho_{\mathcal{A}} \right) ~ .
\end{align}

A generic, random state in the Hilbert space displays an entanglement entropy that is extensive $ S(\rho_{\mathcal{A}}) \sim V_{\mathcal{A}}$. However, states in the MPS subset have an entanglement entropy that is bound by a constant $ S(\rho_{\mathcal{A}}) = \mathcal{O}(\log \chi)$. This is an example of a low entanglement law and is extremely restrictive making matrix product states highly atypical. However, perhaps surprisingly at first, many of the physical states of interest are also low entanglement states. This can be seen somewhat intuitively from the more familiar fact that often physical states are actually quite special in that they do not have arbitrarily long distance correlations. In particular, the ground states of gapped, local systems tend to have exponentially decaying correlations while the ground states of gapless/critical systems tend to have algebraically decaying correlations. As such, one would not expect the entire subregion $\mathcal{A}$ of a system to be correlated with the rest of the system but for the dominant contribution to come from the boundary of the subsystem $\partial\mathcal{A}$. This intuition is indeed correct in many cases and e.g. it has been shown that states with exponentially decaying correlations defined on a ring obey \textit{entanglement area laws} $S({\rho_{\mathcal{A}}}) \sim \partial\mathcal{A} $ \cite{Brandao2015}, see \cite{Eisert2010} for a review of entanglement area laws. 

The understanding that a lot of relevant physics is indeed low entanglement physics has been a driving force behind the development of tensor networks generally, see the review \cite{Eisert2013}. By using a tensor network form to parametrise the Hilbert space of interest, one can truncate down to a tiny low entanglement subset throwing out the vast majority of (highly entangled) states. In this way tensor networks provide a low entanglement effective theory with low entanglement observables being well approximated while high entanglement observables/contributions are lost. In the MPS case, the fact that $ S(\rho_{\mathcal{A}}) = \mathcal{O}(\log \chi)$ means that they are suited to the description of area law states in $D=(1+1)$, since in that case the area law is simply a constant. In fact, one can show that all ground states of local, gapped, lattice Hamiltonians in $D=(1+1)$ can be efficiently represented as MPS and that MPS always have exponentially decaying correlations asymptotically, as expected of an area law state \cite{Eisert2013}. 

In terms of the observables of interest here, the effect of the entanglement cutoff can be seen easily in the connected equal-time two-point function $G_{2}(r) = \braket{\Omega | \phi(x) \phi(x+r) | \Omega} - \braket{\Omega | \phi(x)|\Omega}^{2}$. In the simple lattice truncation, such an observable is approximated by $\braket{\Omega(a,L) | \phi_{x} \phi_{x+r} | \Omega(a,L)}- \braket{\Omega(a,L) | \phi_{x}|\Omega(a,L)}^{2}$ which will agree with the full observable in the region $ a \ll r \ll L$. Outside this region the lattice effects will be significant and the approximation poor. Since the relevant physics of the full theory takes place on a scale determined by the correlation length $\xi$, this means that the physics of the full theory will be well approximated by the lattice theory whenever $ a \ll \xi \ll L$. The additional truncation to the low entanglement subset $\MPS(a,L,d,\chi)$ then restricts this region of validly further and one can think of an additional infrared length scale $\xi_{\chi}$ being introduced after which the inevitable exponential decay of the MPS will dominate and the observable $G_{2}(r)$ will be heavily modified by the entanglement restriction. Additionally, the truncation parameter $\chi$ provides a short distance ultraviolet cutoff which, while essential when studying MPS formulated in the continuum \cite{Haegeman2010}, is made irrelevant by the lattice spacing $a$. Similarly, the lattice size $L$ will tend to be less relevant that the long distance scale $\xi_{\chi}$ and so we can summarise that $G_{2}(r)$ will be well approximated by the MPS theory within the range $ a \ll r \ll \xi_{\chi}$ so that to capture the relevant physics of the full theory we require the hierarchy $ a \ll \xi \ll \xi_{\chi}$.

\subsection{Representation of Observables as Tensor Networks}

\label{Obs}

Once a MPS representation for a state has been obtained, one would like to be able to calculate observables in an efficient way. This can often be achieved by representing the observable as a tensor network. The simplest example  of this is the representation of the overlap between two states $\braket{\tilde{\psi}|\psi} $ which can be found explicitly from the MPS form as
\begin{align}
\braket{\tilde{\psi} | \psi} & = \sum_{\mathbf{n_{x}}} \tilde{\psi}_{\mathbf{n_{x}}}^{*} \psi_{\mathbf{n_{x}}} \nonumber \\
& = \sum_{\mathbf{n_{x}}} \tr \left( \prod_{x}  (\tilde{\M}^{n_{x}})^{*}(x) \right) \tr \left( \prod_{x}  \M^{n_{x}}(x) \right) \nonumber \\
&= \sum_{\mathbf{n_{x}}} \tr \left( \prod_{x}  (\tilde{\M}^{n_{x}})^{*}(x) \otimes \M^{n_{x}}(x) \right) \nonumber \\
&= \tr \left( \prod_{x} \left[ \sum_{n_{x}}   (\tilde{\M}^{n_{x}})^{*}(x) \otimes \M^{n_{x}}(x) \right] \right) .
\end{align}
This expression is clearer in the diagrammatic representation and can be written by introducing the convention that the conjugation of a tensor is represented by flipping the vertical, physical index such that 

\begin{center}

\begin{tikzpicture}
[A/.style = {shape=circle,draw=blue!100,fill=gray!50},node distance = 10pt]
\node (M1) at (0,0) [A] {};
\draw [-] (M1) to (0,0.5);
\draw [-] (M1) -- (-0.5,0);
\draw [-] (M1) -- (0.5,0);
\node (L1) [left= of M1] {$(M^{i}_{\alpha , \beta})^{*} = $};
\node (R1) [right= of M1] {$ . $};

\end{tikzpicture}

\end{center}

The overlap can then be represented as

\begin{center}
\begin{tikzpicture}
[node distance = 10pt ,A/.style = {shape=circle,draw=blue!100,fill=gray!50} , empty/.style = {shape=circle,draw=blue!0},B/.style = {shape=circle,draw=blue!100,fill=gray!0}];

\node[A] (M3) at (0,0) {} ;
\node[A] (M2) [left=of M3] {};
\node[A] (M1) [left=of M2] {};
\node[A] (M4) [right=of M3] {};
\node[A] (M5) [right=of M4] {};

\node[B] (M1B) [below=of M1] {};
\node[B] (M2B) [below=of M2] {};
\node[B] (M3B) [below=of M3] {};
\node[B] (M4B) [below=of M4] {};
\node[B] (M5B) [below=of M5] {};

\draw (M1) -- (M2);
\draw (M2) -- (M3);
\draw ( M3) -- (M4);
\draw (M4) -- (M5);

\draw (M1B) -- (M2B);
\draw (M2B) -- (M3B);
\draw ( M3B) -- (M4B);
\draw (M4B) -- (M5B);

\draw (M1) -- (M1B);
\draw (M2) -- (M2B);
\draw ( M3) -- (M3B);
\draw (M4) -- (M4B);
\draw (M5) -- (M5B);

\node (L1) [left= of M1,yshift=-10pt] {$\braket{\tilde{\psi}|\psi} = $};

\node (R1) [right= of M5,yshift=-10pt] {.};

\end{tikzpicture}

\end{center}
In words, one simply contracts the physical indices of the states together site-by-site. This tensor network representation for the overlap contains no uncontracted indices so that contracting it fully will produce a single number i.e. the value of the overlap. To represent more general operator matrix elements $\braket{\tilde{\psi}|\hat{O}|\psi}$ as tensor networks it is standard to introduce a form for operators that corresponds to the MPS form. Such a representation for lattice systems is called a \textit{matrix product operator} (MPO) form which can be found for a particular operator using some tricks (see Section \ref{RepLH}) or using more generic construction methods \cite{Hubig2017}. To specify an operator, a natural basis choice is the \textit{transition basis} consisting of tensor products of local operators $\hat{T}_{n_{x},m_{x}} = \ket{n_{x}}\bra{m_{x}}$ so that the basis element is $\hat{T}_{\mathbf{n_{x}},\mathbf{m_{x}}} = \hat{T}_{n_{1},m_{1}}\otimes \hat{T}_{n_{2},m_{2}} \otimes ... \otimes  \hat{T}_{n_{L},m_{L}}$ which is labelled by 2 L-tuples $(\mathbf{n_{x}},\mathbf{m_{x}})$. In this basis an operator can be expanded as
\begin{align}
\hat{O} &= \sum_{\mathbf{n_{x}},\mathbf{m_{x}}} \braket{\mathbf{n_{x}}|\hat{O}|\mathbf{m_{x}}} \hat{T}_{\mathbf{n_{x}},\mathbf{m_{x}}} \nonumber \\
&= \sum_{\mathbf{n_{x}},\mathbf{m_{x}}} O_{\mathbf{n_{x}},\mathbf{m_{x}}} \hat{T}_{\mathbf{n_{x}},\mathbf{m_{x}}} ~~ . 
\end{align}
A matrix operator form is then given by rewriting the coefficient $O_{\mathbf{n_{x}},\mathbf{m_{x}}}$ as a set of rank-4 tensors $W_{\alpha_{x},\alpha_{x+1}}^{n_{x} , m_{x}}(x)$ of size $ (d,d,\chi_{W},\chi_{W})$ contracted in nearest-neighbour fashion as
\begin{align}
O_{\mathbf{n_{x}},\mathbf{m_{x}}} &= \sum_{\alpha_{1},\alpha_{2},...,\alpha_{L}} W_{\alpha_{1},\alpha_{2}}^{n_{1} , m_{1}}(1) W_{\alpha_{2},\alpha_{3}}^{n_{2} , m_{2}}(2) ... W_{\alpha_{L},\alpha_{1}}^{n_{L} , m_{L}}(L) \nonumber \\
& = \tr \left( \mathbf{W}^{n_{1} , m_{1}}(1) \mathbf{W}^{n_{2} , m_{2}}(2)  ... \mathbf{W}^{n_{L} , m_{L}}(L) \right) \nonumber \\
& = \tr \left( \prod_{x}\mathbf{W}^{n_{x} , m_{x}}(x)\right)
\end{align}
A diagrammatic expression for operators in (OBC) MPO form is then
\begin{center}
\begin{tikzpicture}
[node distance = 10pt ,A/.style = {shape=circle,draw=blue!100,fill=gray!50} ,O/.style = {shape=rectangle,draw=blue!100,fill=black!50}, empty/.style = {shape=circle,draw=blue!0},B/.style = {shape=circle,draw=blue!100,fill=gray!0}];

\node[empty] (M3) at (0,0) {};
\node[empty] (M2) [left=of M3] {};
\node[empty] (M1) [left=of M2] {};
\node[empty] (M4) [right=of M3] {};
\node[empty] (M5) [right=of M4] {};

\node[O] (O1) [below=of M1] {};
\node[O] (O2) [below=of M2] {};
\node[O] (O3) [below=of M3] {};
\node[O] (O4) [below=of M4] {};
\node[O] (O5) [below=of M5] {}	;

\node[empty] (M1B) [below=of O1] {};
\node[empty] (M2B) [below=of O2] {};
\node[empty] (M3B) [below=of O3] {};
\node[empty] (M4B) [below=of O4] {};
\node[empty] (M5B) [below=of O5] {};

\draw (O1) -- (O2);
\draw (O2) -- (O3);
\draw ( O3) -- (O4);
\draw (O4) -- (O5);

\draw (M1) -- (O1);
\draw (M2) -- (O2);
\draw (M3) -- (O3);
\draw (M4) -- (O4);
\draw (M5) -- (O5);

\draw (O1) -- (M1B);
\draw (O2) -- (M2B);
\draw (O3) -- (M3B);
\draw (O4) -- (M4B);
\draw (O5) -- (M5B);

\node (L1) [left= of O1] {$O_{\mathbf{n_{x}},\mathbf{m_{x}}} = $};

\end{tikzpicture}

\end{center}
such that operator matrix elements can be represented as a tensor network by sandwiching the MPO between two MPS as
\begin{center}
\begin{tikzpicture}
[node distance = 10pt ,A/.style = {shape=circle,draw=blue!100,fill=gray!50} ,O/.style = {shape=rectangle,draw=blue!100,fill=black!50}, empty/.style = {shape=circle,draw=blue!0},B/.style = {shape=circle,draw=blue!100,fill=gray!0}];

\node[A] (M3) at (0,0) {};
\node[A] (M2) [left=of M3] {};
\node[A] (M1) [left=of M2] {};
\node[A] (M4) [right=of M3] {};
\node[A] (M5) [right=of M4] {};

\node[O] (O1) [below=of M1] {};
\node[O] (O2) [below=of M2] {};
\node[O] (O3) [below=of M3] {};
\node[O] (O4) [below=of M4] {};
\node[O] (O5) [below=of M5] {}	;

\node[B] (M1B) [below=of O1] {};
\node[B] (M2B) [below=of O2] {};
\node[B] (M3B) [below=of O3] {};
\node[B] (M4B) [below=of O4] {};
\node[B] (M5B) [below=of O5] {};

\draw (M1) -- (M2);
\draw (M2) -- (M3);
\draw ( M3) -- (M4);
\draw (M4) -- (M5);

\draw (M1B) -- (M2B);
\draw (M2B) -- (M3B);
\draw ( M3B) -- (M4B);
\draw (M4B) -- (M5B);

\draw (O1) -- (O2);
\draw (O2) -- (O3);
\draw ( O3) -- (O4);
\draw (O4) -- (O5);

\draw (M1) -- (O1);
\draw (M2) -- (O2);
\draw (M3) -- (O3);
\draw (M4) -- (O4);
\draw (M5) -- (O5);

\draw (O1) -- (M1B);
\draw (O2) -- (M2B);
\draw (O3) -- (M3B);
\draw (O4) -- (M4B);
\draw (O5) -- (M5B);

\node (L1) [left= of O1] {$\braket{\tilde{\psi}|\hat{O}|\psi} = $};

\node (R1) [right= of O5] {$.$};
\label{ObsTN}
\end{tikzpicture}

\end{center}

Once an observable has been expressed as a tensor network in terms of MPS and MPO, it must be evaluated by contracting the tensors together in the pattern indicated. However, not all patterns will be equally efficient and can be exponentially expensive in the number of sites. In the present case, an efficient contraction ordering is given by proceeding horizontally from the left or right boundary. This can be expressed in terms of transfer matrices by defining the object $E^{A}_{B}[O]$, or $E^{A}_{B}$ if no operator is included, where $A$ and $B$ label the rank-3 tensors placed in the upper and lower positions respectively. Diagrammatically we have

\begin{center}

\begin{tikzpicture}
[node distance = 5pt ,A/.style = {shape=circle,draw=blue!100,fill=gray!50} ,O/.style = {shape=rectangle,draw=blue!100,fill=black!50}, empty/.style = {shape=circle,draw=blue!0},B/.style = {shape=circle,draw=blue!100,fill=gray!0}];

\node[A] (M1)  {};
\node[empty] (O1) [below = of M1] {};
\node[B] (M1B) [below = of O1]  {};

\node[empty] (M1L) [left = of M1] {};
\node[empty] (O1L) [left = of O1] {};
\node[empty] (M1BL) [left = of M1B] {};
\node[empty] (M1R) [right = of M1] {};
\node[empty] (O1R) [right = of O1] {};
\node[empty] (M1BR) [right = of M1B] {};

\draw [-] (M1) -- (M1B);
\draw [-] (M1) -- (M1L);
\draw [-] (M1) -- (M1R);

\draw [-] (M1B) -- (M1BL);
\draw [-] (M1B) -- (M1BR);

\node[empty] (L2) [left=of O1] {$E^{A}_{B} = $};

\end{tikzpicture}

\end{center}
and

\begin{center}

\begin{tikzpicture}
[node distance = 10pt ,A/.style = {shape=circle,draw=blue!100,fill=gray!50} ,O/.style = {shape=rectangle,draw=blue!100,fill=black!50}, empty/.style = {shape=circle,draw=blue!0},B/.style = {shape=circle,draw=blue!100,fill=gray!0}];

\node[A] (M1)  {};
\node[O] (O1) [below = of M1] {};
\node[B] (M1B) [below = of O1]  {};

\node[empty] (M1L) [left = of M1] {};
\node[empty] (O1L) [left = of O1] {};
\node[empty] (M1BL) [left = of M1B] {};
\node[empty] (M1R) [right = of M1] {};
\node[empty] (O1R) [right = of O1] {};
\node[empty] (M1BR) [right = of M1B] {};

\draw [-] (M1) -- (O1);
\draw [-] (O1) -- (M1B);
\draw [-] (M1) -- (M1L);
\draw [-] (M1) -- (M1R);

\draw [-] (O1) -- (O1L);
\draw [-] (O1) -- (O1R);

\draw [-] (M1B) -- (M1BL);
\draw [-] (M1B) -- (M1BR);

\node[empty] (L1) [left=of O1] {$E^{A}_{B}[O] = $};

\node[empty] (R1) [right=of O1] {$.$};

\end{tikzpicture}

\end{center}

In this notation, the matrix element $\braket{\tilde{\psi}|\hat{O}|\psi}$ is given by
\begin{align}
\braket{\tilde{\psi}|\hat{O}|\psi} &= \tr\left( E^{A_{1}}_{B_{1}}[O_{1}] E^{A_{2}}_{B_{2}}[O_{2}] ... E^{A_{L}}_{B_{L}}[O_{L}]\right) \nonumber \\
&= \tr\left( \mathbf{E}_{1} \mathbf{E}_{2} ... \mathbf{E}_{L}\right) \nonumber \\
& = \tr \left( \prod_{x} \mathbf{E}_{x} \right) ~ . 
\end{align}
Viewing $\mathbf{E}_{x}$ as matrix of size $\left(\chi^{2}\chi_{W},\chi^{2}\chi_{W}\right)$ the cost of this matrix multiplication would naively be $\mathcal{O}\left( \chi^{6} \chi_{W}^{3} \right)$ though this can be lowered in a variety of ways which we discuss further in Section \ref{Eff}.

Since any state can be expressed as an MPS and any operator as an MPO this representation for observables can always be used. However, we will only be interested in those observables which can be represented efficiently. For the operators this means finding which can be represented with sufficiently small $\chi_{W}$. The set of operators that can be represented trivially, i.e. with $\chi_{W} = 1$, are simply operators of the tensor product form e.g. all equal-time $n$-point functions. It is not so obvious that other important operators such as the lattice Hamiltonian can also be expressed in an efficient MPO form. However, this is indeed possible and we review the construction in the next section.

\subsection{Representation of Lattice Hamiltonian as MPO}
\label{RepLH}
To represent the lattice Hamiltonian $\tilde{H}[\phi]$ as an MPO it is helpful to first simplify the expression by collecting all one site terms into a single operator $h_{x}$ so that the Hamiltonian takes the form
\begin{align}
\tilde{H}[\phi] = \sum_{x} -\phi_{x}\phi_{x+1} + h_{x} ~
\end{align}
with 
\begin{align}
h_{x} &= \tfrac{2+\tilde{\mu}_{0}^{2}}{2} \phi^{2}_{x} + \tfrac{\tilde{\lambda}_{0}}{4!} \phi^{4}_{x} ~ .
\end{align}
Clearly, this has the same nearest neighbour structure as the Ising model Hamiltonian and we can use the same MPO construction methods as in that case \cite{Wall2012}. A correct MPO representation can then be obtained by building the Hamiltonian up iteratively from the rightmost site to the left. To see this, consider the matrix of operators
\begin{align}
\mathbf{W}_{x} = \begin{pmatrix}
\mathds{1} & 0 & 0 \\ 
-\phi_{x} & 0 & 0 \\ 
h_{x} & \phi_{x} & \mathds{1}
\end{pmatrix}   . 
\end{align}
The matrix is tri-diagonal so that multiplying $\mathbf{W}_{x} \mathbf{W}_{x+1}$ gives
\begin{align}
\mathbf{W}_{x}\mathbf{W}_{x+1} = \begin{pmatrix}
\mathds{1} & 0 & 0 \\ 
-\phi_{x} & 0 & 0 \\ 
h_{x}+h_{x+1}-\phi_{x}\phi_{x+1} & \phi_{x+1} & \mathds{1}
\end{pmatrix}  .
\end{align}
The general structure remains the same while the Hamiltonian is built up in the bottom-left corner. The bulk terms of the Hamiltonian can be built up iteratively in this way. It then only remains to pick a single boundary term (e.g. at site $x=1$) to correctly select out the bottom-left corner, where the bulk of the Hamiltonian has been built up, and encode any remaining boundaries using the trace appearing in the MPO definitions. For PBC a consistent choice is 
\begin{align}
\mathbf{W}_{1} = \begin{pmatrix}
0 & \phi_{1} & \mathds{1} \\ 
0 & 0 & -\phi_{1} \\ 
0 & 0 & h_{1}
\end{pmatrix}
.   
\end{align}
Open boundary conditions can also be encoded by setting $\mathbf{W}_{1}$ and $\mathbf{W}_{L}$ to be vectors $\mathbf{W}_{1} = w_{1} , \mathbf{W}_{L} = w_{L}$  e.g. by the choice
\begin{align}
w_{1}^{T} &= \begin{pmatrix}
h_{1} & \phi_{1} & \mathds{1}
\end{pmatrix} , \\
w_{L} &= \begin{pmatrix}
\mathds{1} \\ -\phi_{L} \\ h_{L}
\end{pmatrix}  .
\end{align}
To specialise to TPBC such that $\phi(x+L) = -\phi(x)$, we can introduce the field variable
\[
 \tilde{\phi}(x) = 
  \begin{cases} 
   \phi(x) & \text{if } x \leq L \\
   -\phi(x)       & \text{if } x > L
  \end{cases}
\]
for $x = 1,2,..,2L$. This means we can treat the finite size Hamiltonian as having an impurity (the twist) at a particular lattice site and observables such as correlations functions must be transformed back to the original variables to keep the periodicity intact. The boundary matrix then takes the form
\begin{align}
\mathbf{W}_{1} = \begin{pmatrix}
0 & -\tilde{\phi}_{1} & \mathds{1} \\ 
0 & 0 & -\tilde{\phi}_{1} \\ 
0 & 0 & h_{1}
\end{pmatrix}   ~ .
\end{align}
Through the above constructions we have found that indeed the nearest-neighbour Hamiltonian $\tilde{H}[\phi]$ can be expressed efficiently as a MPO with $\chi_{W} = 3$, reflecting its local nature. Longer range interactions can also be considered, as would be the case if higher-order finite difference approximations were used. However, in that case larger values of $\chi_{W}$ would be required reflecting the less local nature of the operator. Having an efficient representation of the lattice Hamiltonian as an MPO allows for the efficient computation of observables such as the energy expectation value or energy variance, which are central to obtaining approximations of the ground state and one kink state as MPS.

\section{Approximation of $\phi^{4}$ ground state and one kink state via variational search.}

\label{Min}

\subsection{Variational Energy Minimisation for Matrix Product States}

We now turn to the approximation of the ground state and one kink state using MPS. While the minimisation problems given by Equations (\ref{MinGS}) and (\ref{MinK}) are too difficult to solve in the full lattice Hilbert space, we can restrict them to the MPS subset. The approximations to the states then become
\begin{align}
\ket{\Omega(a,\chi)} = \min_{\ket{\psi} \in \MPS(a,\chi)} \left( \braket{\psi|\tilde{H}|\psi} - \lambda \left[\braket{\psi|\psi} - 1\right]\right) ~
\end{align}
for the ground state and
\begin{align}
\ket{K(a,\chi)} = \min_{\ket{\psi} \in \MPS(a,\chi)} \left( \braket{\psi|\tilde{H}_{\scriptscriptstyle{(TPBC)}}|\psi} - \lambda \left[\braket{\psi|\psi} - 1\right] \right) ~
\end{align}
for the one kink state, where only the most important truncation parameters $a$ and $\chi$ have been kept explicit. Despite the restriction, such a problem is still too hard to solve globally but one can try an iterative procedure that minimises the energy at each step, converging to a best estimate for the global solution. This is now a standard procedure for approximating ground states with MPS and has been highly successful in a variety of cases, see \cite{Schollwock2011} for a detailed guide to implementation. The tensor network structure of the MPS provides a natural way to proceed : one can minimise the energy with respect to just a single tensor (i.e. at a single site) while keeping all other tensors fixed. One then proceeds tensor by tensor minimising the energy iteratively. This is most efficiently performed in a sweeping pattern moving from site to site in a given direction until some convergence criteria are met. In the tensor network representation of observables this local minimisation has a useful form in terms of a generalised eigenvalue problem. To see this, consider the tensor network representation of the energy expectation value with $\tilde{H}$ in MPO form with $L=5$

\begin{center}
\begin{tikzpicture}
[node distance = 10pt ,A/.style = {shape=circle,draw=blue!100,fill=gray!50} ,O/.style = {shape=rectangle,draw=blue!100,fill=black!50}, empty/.style = {shape=circle,draw=blue!0},B/.style = {shape=circle,draw=blue!100,fill=gray!0}];

\node[A] (M3) at (0,0) {};
\node[A] (M2) [left=of M3] {};
\node[A] (M1) [left=of M2] {};
\node[A] (M4) [right=of M3] {};
\node[A] (M5) [right=of M4] {};

\node[O] (O1) [below=of M1] {};
\node[O] (O2) [below=of M2] {};
\node[O] (O3) [below=of M3] {};
\node[O] (O4) [below=of M4] {};
\node[O] (O5) [below=of M5] {}	;

\node[A] (M1B) [below=of O1] {};
\node[A] (M2B) [below=of O2] {};
\node[A] (M3B) [below=of O3] {};
\node[A] (M4B) [below=of O4] {};
\node[A] (M5B) [below=of O5] {};

\draw (M1) -- (M2);
\draw (M2) -- (M3);
\draw ( M3) -- (M4);
\draw (M4) -- (M5);

\draw (M1B) -- (M2B);
\draw (M2B) -- (M3B);
\draw ( M3B) -- (M4B);
\draw (M4B) -- (M5B);

\draw (O1) -- (O2);
\draw (O2) -- (O3);
\draw ( O3) -- (O4);
\draw (O4) -- (O5);

\draw (M1) -- (O1);
\draw (M2) -- (O2);
\draw (M3) -- (O3);
\draw (M4) -- (O4);
\draw (M5) -- (O5);

\draw (O1) -- (M1B);
\draw (O2) -- (M2B);
\draw (O3) -- (M3B);
\draw (O4) -- (M4B);
\draw (O5) -- (M5B);

\node (L1) [left= of O1] {$\braket{\psi|\tilde{H}|\psi} = $};

\node (R1) [right= of O5] {$.$};

\end{tikzpicture}

\end{center}
If we are only interested in varying a single tensor at a particular site, e.g. the central site $x=3$, all other tensors can be contracted together, giving 

\begin{center}
\begin{tikzpicture}
[node distance = 10pt ,A/.style = {shape=circle,draw=blue!100,fill=gray!50} ,O/.style = {shape=rectangle,draw=blue!100,fill=black!50}, empty/.style = {shape=circle,draw=blue!0},OB/.style = {shape=rectangle,draw=blue!100,fill=blue!30}, empty/.style = {shape=circle,draw=blue!0},B/.style = {shape=circle,draw=blue!100,fill=gray!0}];

\node[A] (M1)  {};
\node[O] (O1) [below = of M1] {};
\node[A] (M1B) [below = of O1]  {};

\node [OB] (vL) [left = of O1] {} ;

\draw [-] (O1) -- (O1L);

\node [OB] (v) [right = of O1] {} ;

\draw [-] (O1) -- (M1);
\draw [-] (O1) -- (M1B);

\draw (vL) to (O1);
\draw (vL) to [bend left = 45] (M1) ;
\draw (vL) to [bend right = 45] (M1B);

\draw (v) to (O1);
\draw (v) to [bend right = 45] (M1) ;
\draw (v) to [bend left = 45] (M1B);

\node (L1) [left= of vL] {$\braket{\psi|\tilde{H}|\psi} = $};

\end{tikzpicture}
\end{center}
for OBC while in PBC the boundary tensors would include additional indices to be traced over. This expression can then be thought of as the action of an effective Hamiltonian on the tensor at the uncontracted site. In terms of the tensor network, the effective Hamiltonian then takes the form

\begin{center}
\begin{tikzpicture}
[node distance = 10pt ,A/.style = {shape=circle,draw=blue!100,fill=gray!50} ,O/.style = {shape=rectangle,draw=blue!100,fill=black!50}, empty/.style = {shape=circle,draw=blue!0},OB/.style = {shape=rectangle,draw=blue!100,fill=blue!30},B/.style = {shape=circle,draw=blue!100,fill=gray!0}];

\node[empty] (M1)  {};
\node[O] (O1) [below = of M1] {};
\node[empty] (M1B) [below = of O1]  {};

\node [OB] (vL) [left = of O1] {} ;

\draw [-] (O1) -- (O1L);

\node [OB] (v) [right = of O1] {} ;

\draw [-] (O1) -- (M1);
\draw [-] (O1) -- (M1B);

\draw (vL) to (O1);
\draw (vL) to [bend left = 45] (M1) ;
\draw (vL) to [bend right = 45] (M1B);

\draw (v) to (O1);
\draw (v) to [bend right = 45] (M1) ;
\draw (v) to [bend left = 45] (M1B);

\node (L1) [left= of vL] {$H_{eff} = $};

\end{tikzpicture}
\end{center}
which is a linear operator on the space of rank-3 tensors. As such, it can be considered a matrix of size $(d \chi^{2} , d \chi^{2})$ that acts on vectors  of size $(d \chi^{2})$ (i.e. the rank-3 tensors). In a similar way, an effective normalisation matrix can be constructed by replacing the MPO representing the Hamiltonian by the identity operator via $\braket{\psi | \psi} = \braket{\psi | \mathds{1} | \psi}$ where $ \mathds{1} = \mathds{1}_{1} \otimes \mathds{1}_{2} \otimes ... \otimes \mathds{1}_{L} $. To emphasise this structure we can use the notation $v_{M} = \mathbf{M}^{n_{x}}_{\alpha_{x},\alpha_{x+1}}$ and write the two effective operators as matrices on this space, $\mathbf{H}_{eff}$ and $\mathbf{N}_{eff}$. In this notation the (ground state) minimisation problem at this site can be written as 
\begin{align}
\min_{v_{M} \in \mathds{C}^{d\chi^{2}}} \left( v_{M}^{\dag}\mathbf{H}_{eff}v_{M} - \lambda \left[ v_{M}^{\dag}\mathbf{N}_{eff}v_{M} -1\right] \right) ~ 
\end{align}
which can be solved by finding the minimum eigenvector of the generalised eigenvalue problem 
\begin{align}
\mathbf{H}_{eff}v_{M} = \lambda \mathbf{N}_{eff}v_{M} ~ .
\label{Gen}
\end{align}
To find an approximation to the ground state one then initiates a (random) MPS, chooses a site $i$, forms the effective operators $\mathbf{H}_{eff} , \mathbf{N}_{eff}$, finds the minimum eigenvector $\tilde{v}_{M}$ of the generalised eigenvalue problem Equation (\ref{Gen}) and updates the current MPS by replacing the rank-3 tensor at the site $i$ with the rank-3 tensor corresponding to the minimum eigenvector $\tilde{v}_{M}$ . The updated MPS will have a lower energy and so by proceeding to the next site the energy can be lowered iteratively until convergence is achieved.

The above method can also be used to find an MPS approximation to the one kink state by simply exchanging the PBC Hamiltonian with the TPBC Hamiltonian. However, there are important differences to consider that make the approximation of the one kink state more difficult than the ground state. In particular, the matrix product state techniques described here are naturally inhomogeneous and during the minimisation procedure translational invariance will be broken numerically leading to spatial dependence of the tensors $\M^{n_{x}}(x)$. As such, the translational invariance of observables is only approximated. In the case of the ground state, this is no problem since the MPO representation of the lattice Hamiltonian is reasonably homogeneous and translational invariance can be easily approximated with a low $\chi$ MPS.  However, for the one kink state, the Hamiltonian appears quite inhomogeneous with a particular location being selected for the twist. This makes it much harder to approximate translational invariance and the kink must be ``delocalised" by using a sufficiently high $\chi$. 

The approximation of translational invariance tends to happen quickly so that one can think of a threshold $\tilde{\chi}(d,L)$ after which the spatial variance of local observables drops dramatically. The value of $\tilde{\chi}(d,L)$ will depend on the observable in question as well as the values of $d$ and $L$ with higher $d$ and larger $L$ leading to an increased $\tilde{\chi}(d,L)$. The dependence on $d$ is particularly important since it means that in regions of parameter space requiring high $d$ it will become impossible to approximate translational invariance for the kink state with this method. Since the size of $d$ is determined by the value of the field expectation value $\braket{\phi}$ (or $\braket{\phi^{2}}$ to allow for $\mathds{Z}_{2}$ invariant cases) we see that it is the semi-classical region $\mu_{0}^{2} \gg \lambda_{0}$ that will be hard to approximate in this sense, while the strong coupling region will be less problematic. Of course, this is not too much of an issue since the semi-classical region can be treated perturbatively and observables that include contributions from the entire lattice, e.g. the kink mass, do not depend strongly on the translational invariance of the state. 

Similar issues are present when approximating excited states with MPS more generally and have led to the development of various tensor network excitation ansatz, including one for kinks, that enforce translational invariance explicitly \cite{Haegeman2011}. However, there is some evidence that an inhomogeneous representation could better capture certain aspects of the kinks, which are naturally non-linear field configurations \cite{Berges2011}. This would likely make little difference when considering an observable such as the kink mass, but might be important with observables that have significant contributions from the kink ``width" e.g. the form factor or the equal time two point functions in the presence of a kink near the critical point. We do not explore this further but the excitation ansatz is at least more efficient when determining the mass of excitations near the critical point and we will use them for comparison with the methods here.

\subsection{Computational efficiency, numerical stability and uMPS minimisation}
\label{Eff}

The computational efficiency and numerical stability of the minimisation procedure must be considered at two main stages. Firstly, observables must be calculated efficiently corresponding to the correct choice of contraction ordering. This also covers the construction of the effective operators $H_{eff} , N_{eff}$ which are built up by the partial contraction of a similar tensor network. Secondly, the generalised eigenvalue problem Equation (\ref{Gen}) must be solved. 

In the first case, an efficient contraction pattern is given by simply multiplying the transfer matrices $\mathbf{E}_{x}$ as matrices for a computational cost $\mathcal{O}(\chi^{6})$. A more efficient way to multiply the transfer matrices can be found by making use of their tensor network structure, reducing the cost to $\mathcal{O}(\chi^{4})$. In the case of OBC MPS and MPO, a significant speed up is possible since the boundaries act as vectors and only matrix-vector multiplications are needed. In this case, the naive scaling (i.e. without taking advantage of the tensor network structure) is simply $\mathcal{O}(\chi^{4})$, which can be reduced to $\mathcal{O}(\chi^{3})$ when exploiting the tensor network structure. When considering sufficiently long chains of transfer matrices, the cost of PBC contractions can be reduced as the boundaries become less relevant (becoming completely irrelevant in the infinite distance limit), see \cite{Pippan2010} for details and implementation. In principle this reduces the cost to $\mathcal{O}(\chi^{3})$, equal to OBC, but numerical stability tends to require $\mathcal{O}(\chi^{4})$.

The second case, that of solving the generalised eigenvalue problem, also demonstrates the significant computational advantage of OBC vs PBC. Naively, the cost of solving a generalised eigenvalue problem scales as matrix-matrix multiplication $\mathcal{O}(\chi^{6})$. However, if a sparse implementation is possible then only matrix-vector multiplications are required for a naive cost of $\mathcal{O}(\chi^{4})$. For OBC this can again be reduced to $\mathcal{O}(\chi^{3})$. Unfortunately, solving a generalised eigenvalue problem tends to be ill-conditioned and additional stabilisation steps must be taken. In MPS and tensor networks more generally, stabilisation is often achieved by exploiting the significant \textit{gauge freedom} in the MPS representation. This freedom can be easily seen since any MPS can be equivalently rewritten by inserting identity matrices $\mathds{1}$ of size $(\chi, \chi)$ between any of the rank-$3$ tensors. Decomposing the identities as $\mathds{1} = \mathbf{G}(x)^{-1}\mathbf{G}(x)$ then leads to an equivalent MPS form of the state as
\begin{align}
\psi_{\mathbf{n_{x}}} & =\tr \left( \prod_{x}  \mathbf{G}(x)\M^{n_{x}}(x)\mathbf{G}^{-1}(x+1) \right)  ~ \nonumber \\
& =\tr \left( \prod_{x}  \tilde{\M}^{n_{x}}(x)\right)  ~ . 
\end{align}
Equivalently, and more commonly in practice, one can think of performing a matrix decomposition on the tensors which can be chosen so that the eigenvalue problem to be solved is better conditioned, see \cite{Schollwock2011} for details. In the case of OBC it is possible to choose the gauge such that the effective normalisation matrix simply becomes equal to the identity matrix i.e. the generalised eigenvalue problem is transformed into a standard eigenvalue problem which is considerably more stable. For PBC, this transformation is not possible and we have instead followed the stabilisation strategy outlined in \cite{Pippan2010}. Unfortunately, in the case of TPBC this was not sufficient to be able to solve the generalised eigenvalue problem with sparse methods and we have used dense methods at a cost of $\mathcal{O}(\chi^{6})$. Despite the relative expense, we find that the reachable $\chi \approx 20$ are sufficient for studying the kink mass at strong couplings though for studying other observables e.g. the two point function in the presence of the kink, higher $\chi$ would be needed and an alternative stabilisation strategy would be useful such as the one in \cite{Rossini2011} which was applied to a  spin system with TPBC.

For the approximation of the ground state, since TPBC are not required, it is possible to gain the computational advantage of OBC by using a MPS with an explicitly translationally invariant representation which can be achieved by simply requiring all tensors to be identical i.e. $ \mathbf{M}^{n_{x}}(x) = \mathbf{A}^{n_{x}} $ for all $ x = (1,...,L) $. Since there is no spatial variation this MPS can be defined in the infinite size limit $L \to \infty$ such that the boundaries are irrelevant and the computational savings of OBC can be taken advantage of. Such MPS are called \textit{uniform matrix product state} (uMPS) and, due to the greatly decreased number of free parameters, more standard minimisation procedures can be used to obtain approximations to the ground state \cite{Haegeman2013,Vanderstraeten2015}. Due to the greatly increased efficiency we use uMPS to approximate the ground state following the conjugate gradient procedure outlined in \cite{Milsted2013}. The use of uMPS allows for much higher values of $\chi$ which is essential for capturing the relevant physics of the ground state near the critical point due to the diverging correlation length. Additionally, we note there is an efficient time evolution procedure associated to uMPS known as the \textit{time dependent variational principle} (TDVP) \cite{Haegeman2011} making uMPS a good candidate for the study of non-equilibrium physics. For an open source uMPS code, which aided the development of the code used here, see \cite{Milsted2017}.

\section{Scalar Mass }
\label{Scalar}

\subsection{Long-Distance Behaviour of $G_{2}(r)$}

Since tensor networks represent the state of the quantum system directly, a possible method for obtaining the scalar mass is to try and directly approximate the one particle excitation in the system. However, since tensor networks are often particularly suited to the description of ground states, it is useful to have a general procedure for extracting the scalar mass from ground state observables alone. While it is possible to extract the scalar mass from the long time behaviour of two point functions, for MPS it is much easier to consider the ground state equal time two point functions $G_{2}(r)$ which can be directly calculated once a ground state approximation is found.

To see how the scalar mass can be extracted, we can consider the K\"all\'en-Lehmann spectral representation of the time ordered ground state two point function which can be constructed quite generally for a Lorentz invariant theory \cite{Peskin1995}. This representation relates the full two point function to the two point function of the non-interacting theory, specifically the Feynman propagator $D_{F}(x-y ; M^{2})$ via
\begin{align}
\braket{\Omega|T \phi(x)\phi(y)|\Omega} = \int_{0}^{\infty} \tfrac{dM^{2}}{2 \pi} \rho(M^{2}) D_{F}(x-y;M^{2}) 
\end{align}
where $x$ and $y$ are space-time coordinates, $\rho(M^{2})$ is the spectral density given by
\begin{align}
\rho(M^{2}) &= \sum_{\lambda} (2 \pi) \delta( M^{2} - m_{\lambda}^{2}) |\braket{\Omega|\phi(0)|\lambda_{0}}|^{2}  \\
&= \sum_{\lambda} (2 \pi) \delta( M^{2} - m_{\lambda}^{2}) Z 
\end{align}
and $\ket{\lambda_{0}}$ is a zero-momentum energy eigenstate. We can evaluate the Feynman propagator easily in $D = (1+1)$ at equal times to give
\begin{align}
D_{F}(r ; M^{2})  &= \tfrac{1}{2\pi} K_{0}(M r) ~ 
\end{align}
where $K_{0}(z) $ is a modified Bessel function of the second kind. The equal time two point function can then be written as
\begin{align}
G_{2}(r) = \int_{0}^{\infty} \tfrac{dM^{2}}{4 \pi} \rho(M^{2}) K_{0}(M r) ~ . 
\end{align}
If the spectrum contains an isolated pole then we can extract this contribution and write schematically
\begin{align}
G_{2}(r)  =  \tfrac{Z}{(2 \pi)^{2}} K_{0}(m_{S} r) + \int_{4 M^{2}}^{\infty} \tfrac{dM^{2}}{4 \pi} \rho(M^{2}) K_{0}(M r) ~ . 
\end{align}
This suggests that at sufficiently long distances, the non-interacting form of the two-point function will be dominant and depend on the dimensionless combination $m_{S} r$. 

\subsection{Extracting $m_{S}$ from $G_{2}(r ; a , \chi)$}

The above arguments motivate the use of the ansatz 
\begin{align}
G_{2}(r) = A K_{0}(m_{S} r)
\label{B}
\end{align}
to extract the scalar mass. This can be achieved by taking the appropriate ratios (finite differences) to cancel overall factors as 
\begin{align}
\tfrac{G_{2}(r+1)}{G_{2}(r)}  = \tfrac{K_{0} (m_{s} (r+1))}{K_{0} (m_{s} r)} ~ . 
\end{align}
This equation can then be solved numerically to extract $m_{S}(r)$ which depends on $r$ due the fact that $G_{2}(r)$ is not a pure Bessel function. Following the previous arguments we can expect for some initial $r \lesssim \xi $ the value of $m_{S}(r)$ will vary due to the lattice effects and higher $M^{2}$ eigenstate contributions before becoming uniform such that $m_{S}$ can be extracted (in practice, the uniform region must be selected by some criteria e.g. the gradient of $m_{S}(r)$ falling below some specified tolerance, with $m_{S}$ then estimated by averaging over the selected region). As discussed previously, the use of MPS will modify the long-distance behaviour of the observable $G_{2}(r)$ due to finite entanglement effects ultimately leading to a pure exponential decay. The distance where this occurs is determined by the truncation parameter $\chi$ and we denote the length scale associated to this as $\xi_{\chi}$. We can therefore expect a constant region of $m_{S}(r)$ to occur in some intermediate distance region above the scale of higher mass contributions and below the scale of finite entanglement corrections. A method to diagnose the finite entanglement effects is to repeat the procedure used to extract $m_{S}(r)$ but now use an exponentially decaying form as 
\begin{align}
\tfrac{\delta G_{2} (r)}{ G_{2}(r)} & = \tfrac{\left(G_{2}(r+1)  - G_{2}(r) \right)} { G_{2}(r)} \nonumber \\ 
 &= \tfrac{G_{2}(r+1)}{G_{2}(r)} - 1 \nonumber\\
 &=  e^{-m_{D}} - 1
\end{align}
so that
\begin{align}
m_{D}(r) = -\log( \tfrac{G_{2}(r+1)}{G_{2}(r)}) ~ .
\end{align}
One again, $m_{D}(r)$ will vary with $r$ since only at long distances where the finite entanglement effects dominate will $G_{2}(r)$ become a true exponential. This expression can be used to quickly get a sense of the distance where the correlation function becomes strongly modified by the finite entanglement effects of the MPS. 

\subsection{Extracting $m_{S}$ for Lattice MPS at Strong Couplings}

At strong couplings, the lightest excitations will become kink-antikink pairs. This motivates the use of a Bessel-squared function ansatz for the two point function 
\begin{align}
G_{2}(r) = A K_{0}(m_{S} r)^{2}
\label{B2}
\end{align}
corresponding to the form for two non-interacting excitations. The behaviour of $G_{2}(r)$ can be established more rigorously in the critical region by considering the critical behaviour of the classical $D=2$ Ising model which is described by a field theory of free massive Majorana fermions \cite{Yurov1991}. 

At strong couplings, the Bessel-squared ansatz can then be used in a similar way to the Bessel function form. In fact, both $K_{0} (z)$ and $K_{0} (z)^{2}$ have similar exponentially decaying asymptotic forms 
\begin{align}
K_{0} (z) &\to e^{-z}\sqrt{\tfrac{1}{z}} \\
[K_{0} (z)]^{2} &\to e^{-2 z}\tfrac{1}{2z} ~ . 
\end{align}
As such, in principle either form can be used to estimate $m_{S}$. However, unlike $K_{0} (z)$, we can expect $[K_{0} (z)]^{2}$ to be valid outside the asymptotic regime which is essential since then shorter distances of $G_{2}(r)$ will need to be approximated, requiring smaller $\chi$.

\section{Results}
\label{Res}

\begin{figure*}[t]
\centering
\includegraphics[width=1.0\linewidth]{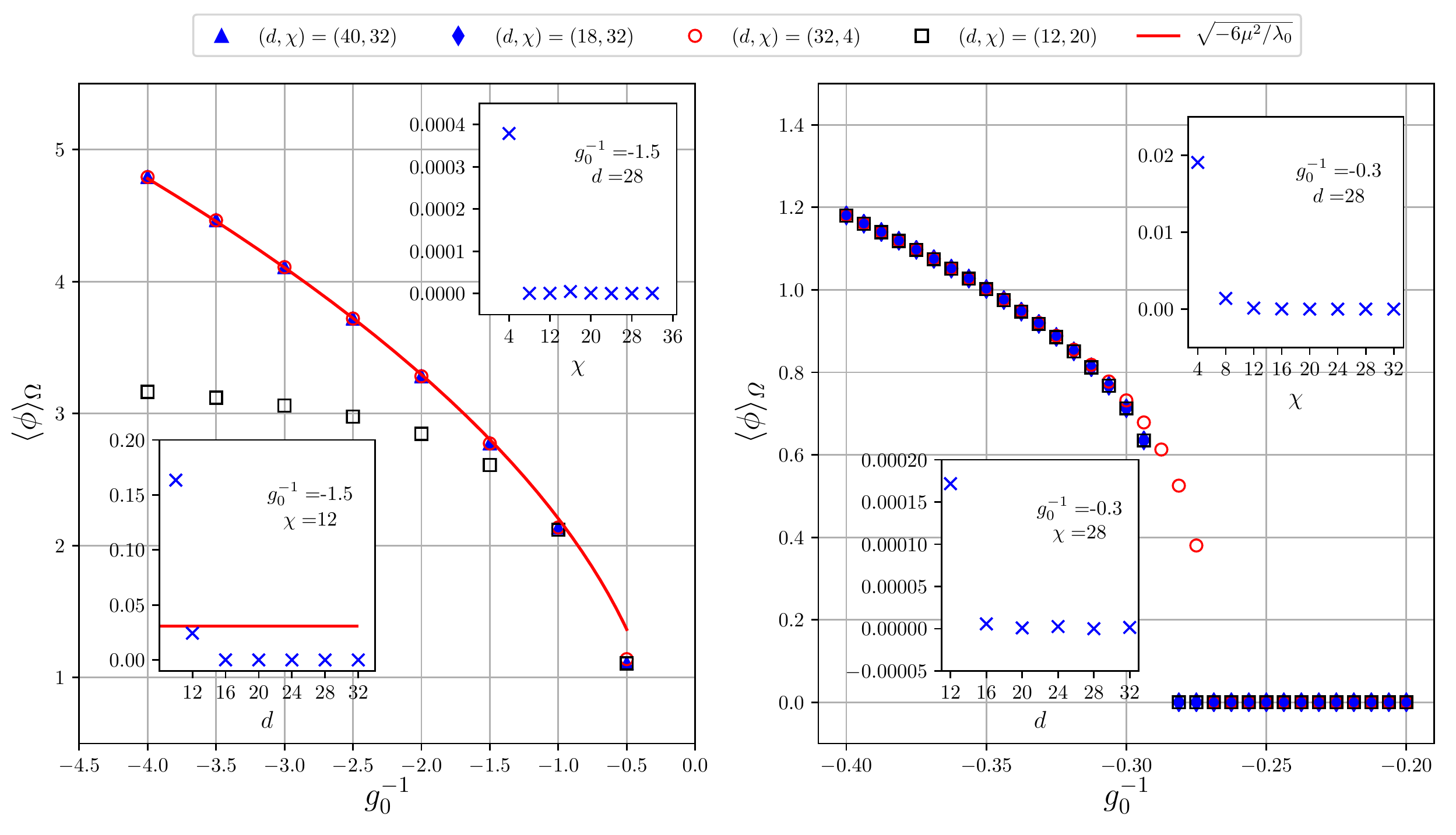}
\caption{The field expectation value for various effective couplings using a uMPS approximation to the ground state. In the left-hand plot, weak-couplings are displayed with $\lambda_{0} = 0.1$ and $(d,\chi) = (40,32) $ (blue triangles). These can be compared to low $\chi = 4$ approximations (hollow red circles) which look almost identical on this scale. This contrasts the low $d = 12$ approximations (hollow black squares) for which the expectation value appears truncated agreeing with the higher $d$ approximations only when the field expectation value is relatively low. The insets show the convergence for the selected inverse coupling $g_{0}^{-1} = -1.5$ (the best estimate has been subtracted so that the plots tend to zero) and the difference in scales between the two indicate that $d$ is the relevant parameter to achieve a good approximation in the weak coupling regime. The semi-classical result is also shown in the main figure and lower-left inset (solid red line) with $ v = \sqrt{-6\mu^{2}/\lambda_{0}}$ where $\mu^{2} = \mu_{0}^{2} - m_{C}^{2}$ and $m_{C}^{2} = -0.019$ is determined by fitting to the data.  The stronger coupling data (right-hand plot) with $(d,\chi) = (18,32)$ and $\lambda_{0} = 2$ (blue diamonds) shows the usual symmetry breaking pattern. Here, the relevant parameter to achieve a good approximation is $\chi$ as shown by the scale difference between the insets and also by the fact that the low $d$ approximation agrees well with the data on the main plot while the low $\chi$ data fails to agree near the critical point.}
\label{VevWCSC}
\end{figure*}

We study the $\phi^{4}$ QFT using MPS in the lattice regularised setting. An approximation to the ground state is obtained using uMPS while a finite size lattice MPS is used to approximate the one kink state. We study both weak and strong coupling behaviour by fixing the value of $\lambda_{0} = 0.1$ or $2$ while using a range of $\mu_{0}^{2}$. When the effective coupling is small we compare the MPS approximation to the classical continuum. The lowest order mass renormalisation is included numerically by fitting the classical forms to the MPS data, replacing the bare mass with $\mu^{2} = \mu_{0}^{2} - m_{C}^{2}$ where $m_{C}^{2}$ is treated as a free parameter. In the region $\xi \gg 1$ the lattice effects will be small and the comparison with the classical continuum is appropriate. In the strong coupling region the MPS approximations can be compared with universal results and we will focus on a comparison of the mass ratio $m_{S}/M_{K}$ to the universal result $m_{S}/M_{K} \approx 2$. 

The accuracy of our approximations will depend on the observable in question. Essentially, the important features are the observation distance and to what degree the observable represents an average over the system. In the first case, only the truncation parameter $\chi$ is important and increasing $\chi$ will allow longer distances to be better approximated. This can be seen clearly in ground state connected two point function $G_{2}(r)$ where larger $\chi$ are required to approximate the observable at larger distances $r$. We can associate this behaviour to a length scale $\xi_{\chi}$ corresponding to the distance at which the approximation of $G_{2}(r)$ is dominated by finite entanglement effects and decays as a pure exponential. In the second case, one can think of a particular threshold $ \chi \approx \tilde{\chi}(d,L)$ being required before the translational invariance of observables is well approximated. This will depend strongly on the observable/state in question and on the truncation parameters $d$ and $L$.

\begin{figure*}[t]
\centering
\includegraphics[width=1.0\linewidth]{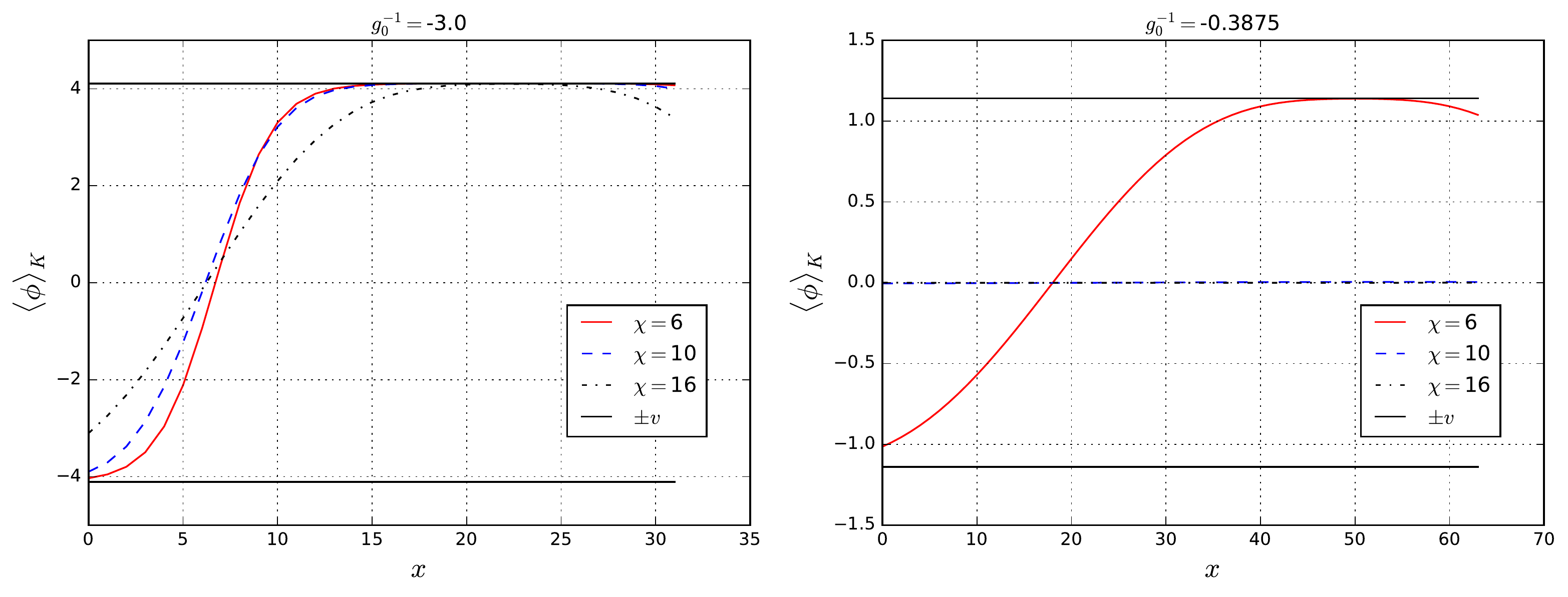}
\caption{The field expectation value of the finite size lattice MPS approximation to the one kink state for weak coupling $ g_{0} \approx -0.33$ (left-hand plot) and stronger coupling $ g_{0} \approx -2.58 $ (right-hand plot) with lattice sizes $L = 32 , 64$ and $ 28 \le d\le 32 ,  14 \le d \le 18$ respectively. These can be compared with the corresponding field expectation values of a uMPS approximation to the ground state $\pm v$ (solid black lines) with $(d,\chi) = (40,32)$ and $(18,32)$ for the weak coupling and stronger coupling respectively. For the low $\chi = 6$ runs (solid red line) a classical kink-like profile is visible for both couplings which interpolates between $\pm v$ such that the correct $\braket{\phi}_{K} = 0$ is not found. Increasing $\chi$ de-localises the kink and translational invariance can be approximated in the stronger coupling case such that $\braket{\phi}_{K} \approx 0$ .}
\label{VevTWCSC}
\end{figure*}

The issues surrounding the approximation of translational invariance can be seen clearly when calculating the field expectation values $\braket{\Omega|\phi|\Omega}$ and $\braket{K|\phi|K}$. Since both observables are local, they will converge quickly in $\chi$ so long as the threshold $\chi \approx \tilde{\chi}(d,L)$ is met. In the ground state case, $\tilde{\chi}(d,L)$ is essentially negligible and the approximation of $\braket{\Omega|\phi|\Omega}$ converges rapidly. However, in the one kink state $\tilde{\chi}(d,L)$ is important and, for sufficiently large values of $d$ or $L$, local expectation values such as $\braket{K|\phi|K}$ will show significant spatial variations and cannot be accurately approximated. We note that, even in the ground-state case, the field expectation value is in princible zero, respecting the $\mathds{Z}_{2}$ symmetry but is broken numerically during the approximation and must be enforced explicitly if desired \cite{Singh2013}. The uMPS approximation of the vacuum expectation value $\braket{\Omega|\phi|\Omega}$ is shown in Figure \ref{VevWCSC} for perturbative and non-perturbative bare couplings $g_{0} = \lambda_{0}/\mu_{0}^{2}$. In the first case, the results (blue triangles) can be compared with the classical continuum result (red line) $ v = \sqrt{-6\mu^{2}/\lambda_{0}}$. In the stronger coupling case, there is no analytic comparison but the expected symmetry breaking pattern can be seen. In principle, such plots can be used to determine the location of the critical point e.g. by using the critical exponent associated with the vanishing field expectation value. However, as discussed in \cite{Milsted2013}, such fits are highly sensitive and it is much better to use observables with a simpler scaling, e.g. the kink mass, to determine the location of the critical point.

The more problematic behaviour of $\braket{K|\phi|K}$ is shown in Figure \ref{VevTWCSC}. In the semi-classical case with bare coupling $g_{0} \approx -0.33$, the high field expectation value requires a relatively high value of $d$ to converge and the threshold $\tilde{\chi}(d,L)$ is higher than the shown $\chi = 6, 10, 16$. This means that a classical-like kink profile can be seen and increasing $\chi$ achieves only very slight changes to the width such that the correct $\braket{\phi}_{K} = 0$ value is not obtained. Moreover, the zero-mode means that the point at which $\braket{\phi(x)}_{K}$ crosses zero is independent of the energy making convergence in $\chi$ or $d$ difficult to quantify.  However, at stronger couplings the field expectation value is much lower, corresponding to a lower $d$, which makes it easy to approximate translational invariance and obtain $\braket{\phi}_{K} \approx 0$ even for the modest values of $\chi$ shown. 

\begin{figure*}[t]
\centering
\includegraphics[width=1.0\linewidth]{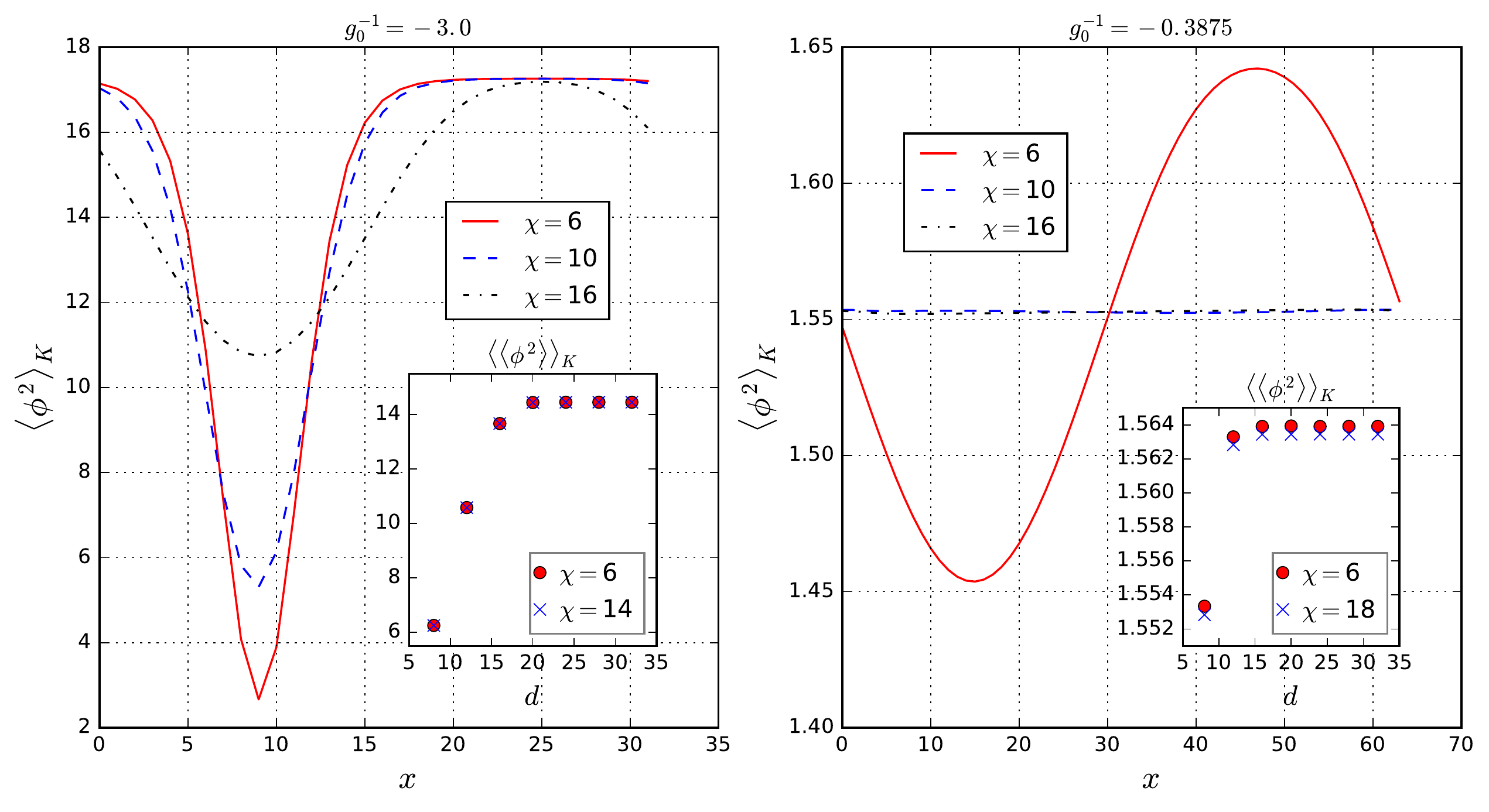}

\caption{The $\phi^{2}$ expectation value corresponding to the one kink approximations in Figure \ref{VevTWCSC}. As with $\braket{\phi}_{K}$, translational invariance can only be approximated in the stronger coupling case.  Nevertheless, since the operator is $\mathds{Z}_{2}$ invariant, the spatial average is well behaved and its convergence can be studied as shown in the insets.}

 \label{Vev2TWCSC}
\end{figure*}

\begin{figure*}[t]
\centering
\includegraphics[width=1.0\linewidth]{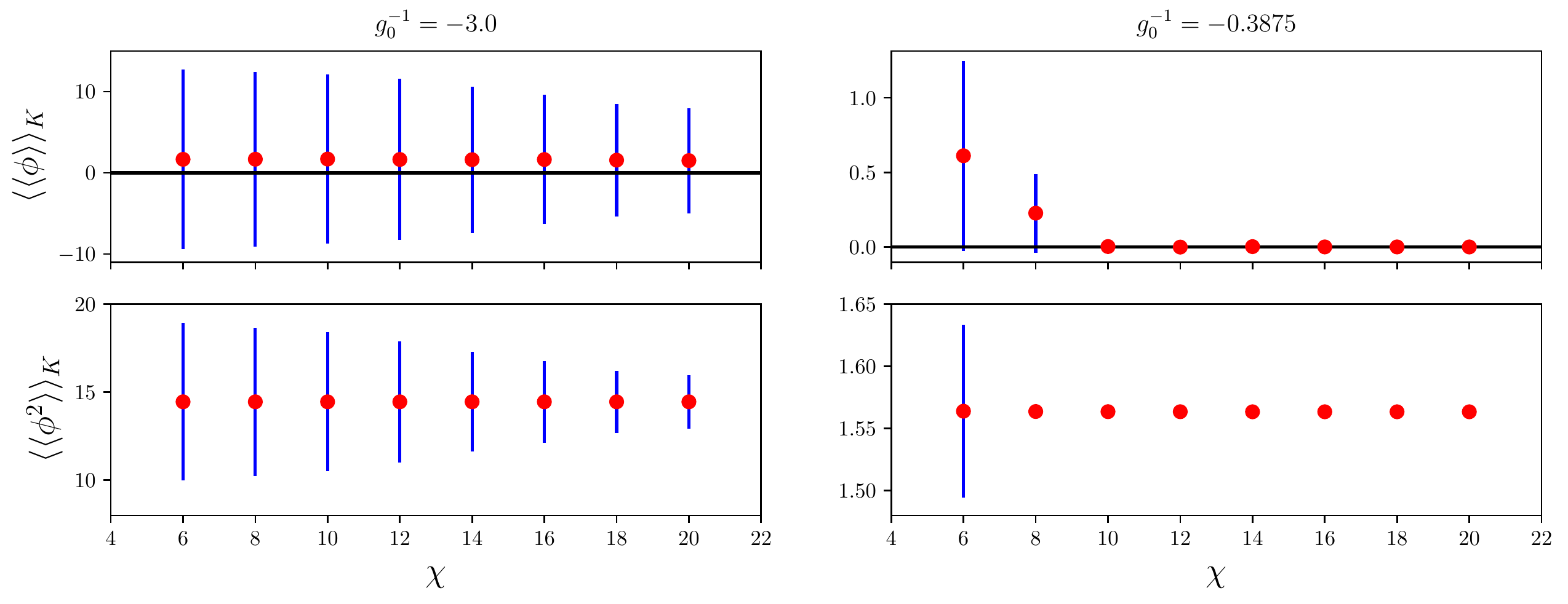}
\caption{The spatial average (red dots) of the $\phi$ and $\phi^{2}$ expectation values for the finite size lattice MPS approximation to the one kink state at weak coupling $ g_{0} \approx -0.33$ (left-hand plots) and stronger coupling $ g_{0} \approx -2.58 $ (right-hand plots) with the same parameters as in Figure \ref{VevTWCSC}. The spatial standard deviation of the expectation values, shown as error bars, changes strongly with $\chi$ and becomes small only for strong couplings. However, the spatial averages changes only weakly and the value of $\braket{\braket{\phi^{2}}}_{K}$ can be reliably approximated in both regions. Note that, due to its anti-symmetry, the value of $\braket{\braket{\phi}}_{K}$ is only correct and equal to zero (indicated by the black line) in the stronger coupling case.}
\label{aVev2TWCSC}
\end{figure*}

Observables that average over the whole system can be much less sensitive to spatial variations in the MPS representation than observables evaluated at a particular point. For example, the behaviour of $\braket{K|\phi^{2}(x)|K} = \braket{\phi^{2}}_{K}$ displays similar spatial variations at weak coupling as for the case of $\braket{K|\phi(x)|K}$, see Figure \ref{Vev2TWCSC} . However, the spatial average of this expectation value $\braket{\braket{\phi^{2}}}_{K}$ has a much weaker dependence on $\chi$. This is shown in Figure \ref{aVev2TWCSC} where the spatial variation of the expectation values of both $\braket{\braket{\phi}}_{K}$ and $\braket{\braket{\phi^{2}}}_{K}$ are shown by error bars corresponding to their standard deviation with $x$. Despite the strong spatial variation in $\braket{\phi^{2}}_{K}$, the spatial average changes only very weakly with $\chi$ in both cases indicating that this observable can be well approximated even in the weak coupling case. This behaviour can be compared with that of $\braket{\phi}_{K}$ ; while the spatial average does not display much variation in the weak coupling case, since the operator is $\mathds{Z}_{2}$ anti-symmetric, it still gives the incorrect non-zero value and is only correctly approximated in the stronger coupling region where translational invariance is approximated. In general, observables corresponding to $\mathds{Z}_{2}$ anti-symmetric operators cannot be reliably approximated outside the translational invariant region, while the spatial average of those corresponding to $\mathds{Z}_{2}$ symmetric operators can be. 

\begin{figure*}[t]
\centering
\includegraphics[width=1.0\linewidth]{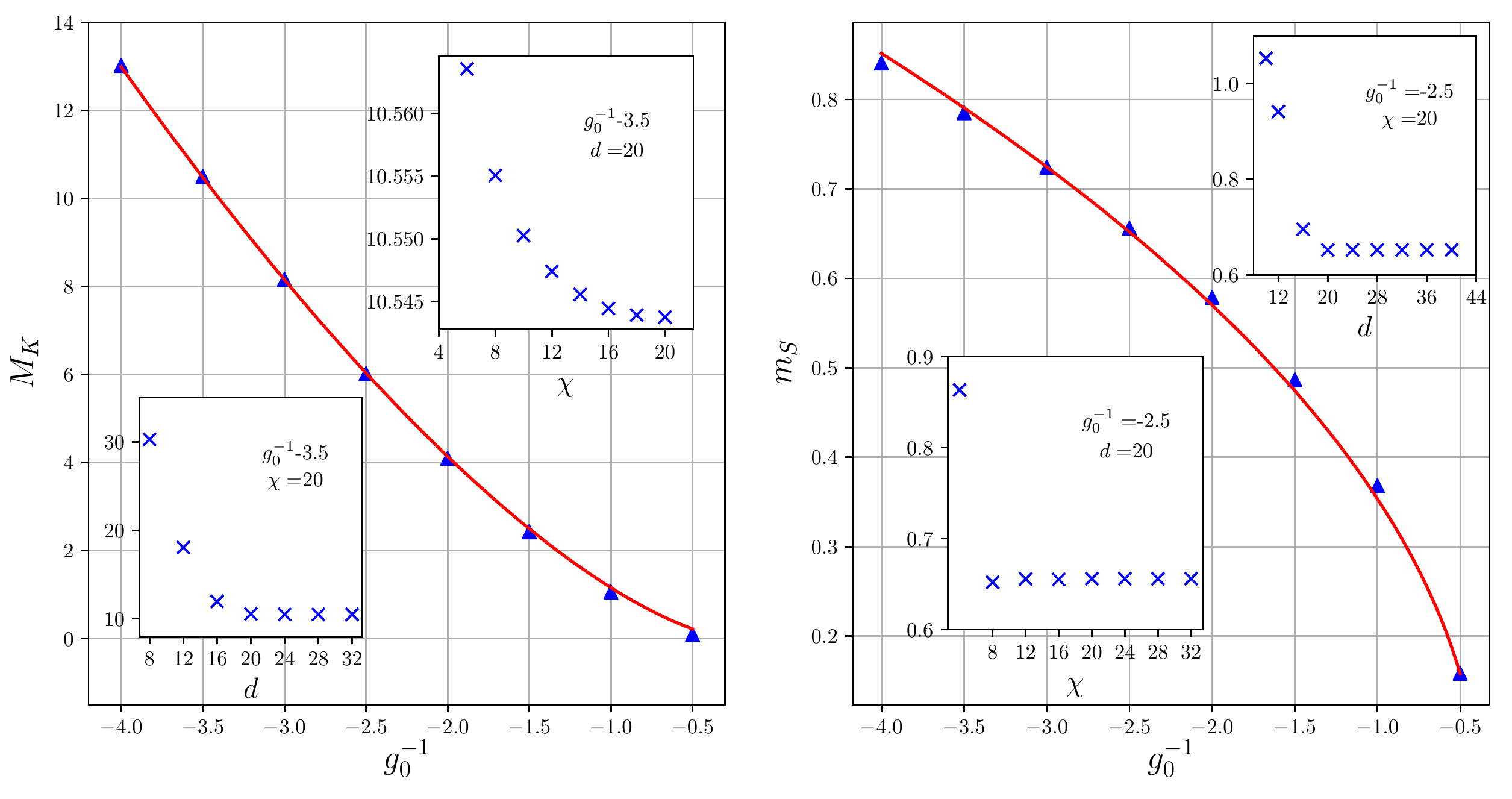}
\caption{The kink mass $M_{K}$ (left-hand plot) and scalar mass $m_{S}$ (right-hand plot) for various weak couplings (data in blue triangles). The kink mass is calculated from the energy expectation value of the finite size lattice MPS approximation of the one kink state with $\chi = 14$ and the approximation of the ground state energy density obtained from a uMPS approximation with $\chi = 32$. The scalar mass is extracted from the uMPS approximation to the ground state connected equal time two point function $G_{2}(r)$ via a Bessel function ansatz (\ref{B}). Both are compared with the semi-classical continuum results (solid red lines) $M_{K} = 4 \sqrt{2} \mu^{3}/\lambda_{0} $ and $m_{S} = \sqrt{2} \mu$ where $\mu^{2} = \mu_{0}^{2} - m_{C}^{2}$ and $m_{C}^{2}$ is determined by fitting to the data to give $ -0.025 , -0.037 $ respectively. The convergence of the approximations with $d$ and $\chi$ is shown in the insets.}
\label{MkSM_WC}
\end{figure*}

The kink mass $M_{K}$ is calculated from the difference of the one kink energy expectation value $\braket{K|\tilde{H}_{TPBC}|K}$, obtained from the finite size lattice MPS, and the ground state energy density, obtained from the uMPS. The latter converges quickly in $\chi$ and the only potential issue is the approximation of $\braket{K|\tilde{H}_{TPBC}|K}$. However, since the kink mass both includes contributions from the whole system, is fairly local and $\mathds{Z}_{2}$ symmetric, we can expect a reasonable convergence with $\chi$ even in the weak coupling case. In the case of the scalar mass, since it is estimated from the ground state observable $G_{2}(r)$, approximating translational invariance should not be an issue and we can focus on the need to increase $\chi$ so that the region up to $r \approx \xi$ is well approximated. At weak couplings, $\xi $ is relatively small so that the required $\chi$ should not be too high allowing for an estimate of $m_{S}$ to be extracted relatively easily. The kink mass and scalar mass are shown for a variety of weak couplings in Figure \ref{MkSM_WC} along with the classical continuum results for comparison. 

At stronger couplings the correlation length $\xi$ increases so that longer distances of $G_{2}(r)$ need to be approximated requiring larger $\chi$. Ultimately, this means that this method cannot be used with MPS arbitrarily close to the critical point where the scalar mass vanishes. This is reflected in the fact that at the critical point the correlation length diverges leading to algebraically decaying correlations which correspond to a logarithmic violation of the entanglement area law i.e. $S_{\mathcal{A}} \sim \log(L_{\mathcal{A}}) \partial \mathcal{A}$. While an MPS can still be used to approximate short distance observables in the critical region \cite{Pirvu2012} an alternative tensor network , e.g. the \textit{multi-scale entanglement renormalisation ansatz} (MERA) \cite{Vidal2006}, that obeys the correct low entanglement law will also allow the approximation of the long range physics. Of course, this means that MPS are not especially suited to the study of universal physics and we can expect difficulty when trying to reproduce the strong coupling behaviour e.g. the universal mass ratio.

\begin{figure*}[t]
\centering
\includegraphics[width=1.0\linewidth]{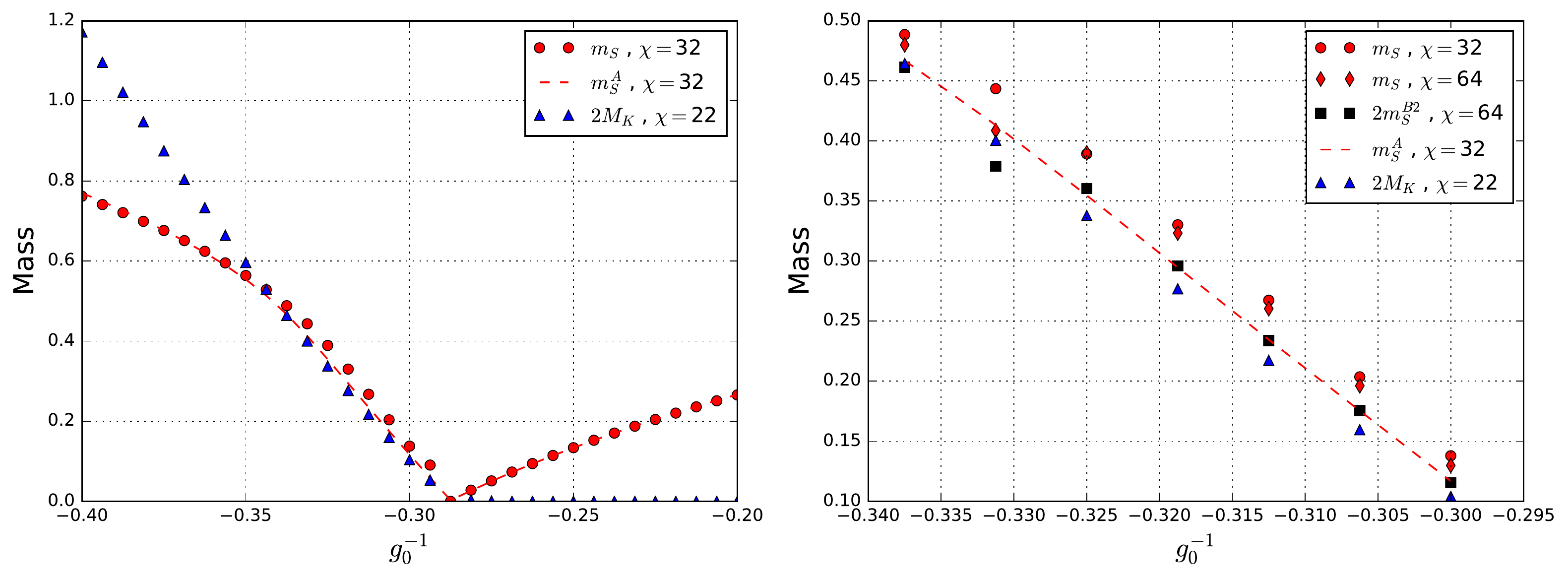}
\caption{Estimate of the scalar mass $m_{S}$ extracted from the uMPS approximation to the ground state connected equal time two point function $G_{2}(r)$ via a Bessel function ansatz (\ref{B}) ($\chi = 32 $ red dots , $\chi = 64$ red diamonds). This can be compared with the estimate extracted from the excitation ansatz $m_{S}^A$ (dashed red line) and twice the kink mass $2 M_{K}$ (blue triangles) calculated from the $L=64$ finite size lattice MPS approximation to the one kink state. The left-hand plot shows a larger range of bare coupling $ 2.5 \lesssim g_{0} \lesssim 5$ while the right-hand plot focuses on the strong coupling region with $2.95 \lesssim g_{0} \lesssim 3.39$. A qualitative change can be seen when entering the strong coupling region at $m_{S} \approx 2 M_{K}$ but the $\chi = 32$ Bessel ansatz (\ref{B}) does not provide a good quantitative agreement with the expected behaviour $m_{S} \approx 2 M_{K}$ in the strong coupling region.  The higher $\chi = 64$ Bessel ansatz does improve the estimate but the $\chi = 64$ Bessel-squared ansatz (\ref{B2}) (black squares, only in right-hand plot) improves the estimate further in agreement with the $\chi = 32$ excitation ansatz estimate (dashed red line). As mentioned in Section \ref{Scalar}, the scalar mass is estimated from averaging over a ``uniform" region of $m_{S}(r)$ chosen here by a single tolerance for all values of $g_{0}^{-1}$. While this has the advantage of being ``blind" using a single tolerance can lead to somewhat anomalous points (e.g. see the point at $g_{0}^{-1} \approx -0.331$) and instead one can choose the tolerance adaptively for each $g_{0}^{-1}$ which eliminates such points, improving the estimates of $m_{S}$.}
\label{MkSM_SC12}
\end{figure*}

The scalar mass at strong couplings is plotted along with the kink mass in Figure \ref{MkSM_SC12} and the qualitative change in the scaling can been seen in the left-hand plot at the point when $2 M_{K} \approx m_{S}$ as expected. However, the estimate of $m_{S}$ with $\chi = 32$ in the critical region extracted from the Bessel function tends to be somewhat higher than the value of $2 M_{K}$ suggesting that, as might be expected, it is inaccurate in this region. The scalar mass extracted from the excitation ansatz with $\chi = 32$ is also plotted (red dashed line) and is somewhat closer to the value of $2 M_{K}$ suggesting that it can provide a more efficient and accurate method to extract the scalar mass in the critical region. To increase the accuracy of the uMPS method one can simply increase the value of $\chi$ but it is also possible to use the Bessel-squared ansatz Equation (\ref{B2}). A comparison of these methods is shown for the strong coupling region in the right-hand plot. The estimate of the scalar mass is closer to the expected behaviour when $\chi$ is increased (red dots and diamonds) but the use of the Bessel-squared ansatz improves the estimate again (black squares) agreeing fairly well with the excitation ansatz. The significant improvement of the Bessel-squared method over the single Bessel method suggests that the uMPS is able to capture the contributions coming from the kink-antikink excitations in this observable. To achieve higher accuracies than obtained here, larger $\chi$ can be used by following more recently developed algorithms than the conjugate gradient minimisation used here  \cite{Vanderstraeten2015}. Alternatively, one can also turn to better suited tensor networks such as MERA and both the methods to obtain the kink mass and scalar mass should be easily adaptable to this case. Of course, if one is only interested in the equilibrium physics chosen here, the excitation ansatz provides good accuracy and efficiency. However, it is less flexible and cannot be readily applied to different tensor networks and instead must be built explicitly for each case. 

\section{Conclusion}
\label{Concl}

We have studied the topological defects (kinks) of the relativistic $\phi^{4}$ quantum field theory (QFT) in $D=(1+1)$ using matrix product states (MPS). We have shown how a finite size lattice MPS approximation to the one kink state can be obtained by making use of twisted periodic boundary conditions (TPBC) and that the resulting kink mass agrees with expectations. While alternative specific excitation ansatz provide a more efficient method to calculate the kink mass, the TPBC method can be easily adapted to other theories and tensor networks while also allowing for the easy calculation of a wide variety of observables e.g. equal time field n-point functions in the presence of the kink. We have also outlined a general method to extract the scalar mass from the ground state equal time two point functions. A comparison of the one kink and ground state approximations with universal results suggests that the MPS (specifically the uMPS) is able to capture the contribution of kink-antikink excitations to observables making it an interesting candidate to study challenging non-equilibrium phenomena such as defect formation via the Kibble Zurek mechanism during quantum phase transitions.
\section*{Acknowledgements}

We have made use of the Imperial College London High Performance Computing Service. E.G. was supported by the EPSRC Centre for Doctoral Training in Controlled Quantum Dynamics and A.R. by STFC grant ST/L00044X/1.

\bibliography{TD_MPS}
\end{document}